\documentclass[a4paper, amsfonts, amssymb, amsmath, reprint, showkeys, nofootinbib, twoside,notitlepage,onecolumn]{revtex4-2}

\def\prd{{Phys.~Rev.~D}}        
\def\prl{{Phys.~Rev.~Lett.}}    
\usepackage{amsmath,amstext}
\usepackage[T1]{fontenc}
\usepackage{amssymb}
\input{epsf}
\usepackage{graphicx}
\usepackage{ae,aecompl}

\usepackage{hyperref}
\usepackage{amsmath}
\usepackage{amssymb}
\usepackage{mathtools}
\usepackage{bm}
\usepackage{cleveref}
\usepackage{tensor}
\usepackage{braket}
\usepackage{enumitem}
\usepackage{mhchem}
\usepackage{cancel}
\usepackage{amsthm}
\usepackage{upgreek}
\usepackage{mathrsfs}
\usepackage{pgfplots}
\usepackage{xcolor}
\usepackage{soul}
\usepackage[normalem]{ulem}

\usepackage{color}

\newcommand{\spc}{\quad \quad \quad}

\newcommand{\Da}{{\mathbb{D}}}
\newcommand{\Ma}{{\mathbb{M}}}
\newcommand{\Sa}{{\mathbb{S}}}
\newcommand{\Ga}{{\mathbb{G}}}
\newcommand{\Qa}{{\mathbb{Q}}}
\newcommand{\Pa}{{\mathbb{P}}}
\newcommand{\Ia}{{\mathbb{I}}}
\newcommand{\Ka}{{\mathbb{K}}}
\newcommand{\La}{{\mathbb{L}}}
\newcommand{\Ha}{{\mathbb{H}}}
\newcommand{\x}{{\textbf{x}}}
\newcommand{\y}{{\textbf{y}}}
\newcommand{\kb}{{\textbf{k}}}

\def\be{\begin{equation}}
\def\ee{\end{equation}}
\def\beq{\begin{eqnarray}}
\def\eeq{\end{eqnarray}}

\theoremstyle{definition}

\begin{document}
\title{Consistent inclusion of fluctuations in first-order causal and stable relativistic hydrodynamics}
\author{ L.~Gavassino$^1$, N.~Mullins$^2$, M.~Hippert$^2$ }
\affiliation{
$^1$Department of Mathematics, Vanderbilt University, Nashville, TN, USA
\\
$^2$Illinois Center for Advanced Studies of the Universe \& Department of Physics,
University of Illinois at Urbana-Champaign, Urbana, IL 61801-3003, USA
}

\begin{abstract}
We construct, for the first time, a BDNK theory for linear stochastic fluctuations, which is proved to be mathematically consistent, causal, and covariantly stable. The Martin-Siggia-Rose action is shown to be bilocal in most cases, and the noise is not white. The presence of non-hydrodynamic modes induces long-range correlations in the primary fluid variables (temperature, chemical potential, and flow velocity). However, correlators of conserved densities remain localized in space, and coincide with those calculated within fluctuating Isreal-Stewart theory. We show that, in some cases, there is a non-local change of variables that maps the Israel-Stewart action into the BDNK action.
\end{abstract}

\maketitle

\section{Introduction}

Every system that dissipates must also fluctuate. This is a universal law of statistical mechanics, known as the \textit{fluctuation-dissipation theorem} (FDT) \cite{landau5,Pathria2011}, whose physical content can be summarised with the following simple argument. 
Consider an isolated system, and let $X$ be a list of observables that we use to characterize its macroscopic state. In a fluid, these are the hydrodynamic fields. We call $\mathcal{P}_e=\mathcal{P}(|X{-}\langle X \rangle|\leq \epsilon)$ the probability of finding $X$ within a distance $\epsilon$ from the microcanonical average $\langle X \rangle$, and $\mathcal{P}_n=1-\mathcal{P}_e$ the probability for the complementary event. The evolution equation of these two probabilities can be modeled schematically as follows:
\begin{equation}\label{processo}
\begin{bmatrix}
 \mathcal{P}_e(t{+}\Delta t) \\
 \mathcal{P}_n(t{+}\Delta t) \\
\end{bmatrix}
=
\begin{bmatrix}
  1{-}\mathcal{P}_{e\rightarrow n}  &  \mathcal{P}_{n\rightarrow e} \\
  \mathcal{P}_{e\rightarrow n} &   1{-}\mathcal{P}_{n\rightarrow e} \\
\end{bmatrix}
\begin{bmatrix}
 \mathcal{P}_e(t) \\
 \mathcal{P}_n(t) \\
\end{bmatrix} \, ,
\end{equation}
where $\mathcal{P}_{i\rightarrow j}$ is the probability of spontaneously jumping from the event $i$ to the event $j$ in a time $\Delta t$. Since the microcanonical probability $\mathcal{P}_i\propto e^{S_i}$ ($=$ ``number of microscopic realizations of $i$'') is stationary under time evolution, it must be conserved under the process \eqref{processo}, giving (a schematic representation of) the FDT\footnote{In the continuous version of \eqref{FDTGeneral}, one assumes Fokker-Planck dynamics, $\partial_t \mathcal{P}{=}{-}\partial_X(F\mathcal{P}){+}\partial^2_X(Q\mathcal{P})$, for the probability distribution $\mathcal{P}(X)$, with dissipative force $F(X)$ and noise scale $Q$. Requiring $e^{S(X)}$ to be a stationary solution, one gets $Q=(\partial_X S)^{-1} \, F$, so that a larger dissipative drag $F$ corresponds to a larger noise scale $Q$, at fixed entropy. See \cite{Kubo:1966fyg} for more general formulations of the FDT, or \cite{Calzetta:1997aj} for a formulation in a relativistic context.}: 
\begin{equation}\label{FDTGeneral}
\mathcal{P}_{e\rightarrow n} = e^{S_n-S_e} 
 \, \mathcal{P}_{n \rightarrow e} \, .
\end{equation}
Hence, if there is a dissipative mechanism $\mathcal{P}_{n\rightarrow e}$ that drives $X$ toward its equilibrium average $\langle X \rangle$, then there must also be a random force $\mathcal{P}_{e\rightarrow n}$ that spontaneously pushes $X$ away from it. The rates of the two processes are proportional to each other, meaning that, fixed the entropies $S_i$, increasing dissipation leads to a corresponding increase in fluctuations.

In everyday applications of hydrodynamics, the effect of fluctuations is often negligible \cite{huang_book}. This is due to the factor $e^{S_n-S_e}$ in the FDT \eqref{FDTGeneral}, which is usually minute, as it scales like $\sim e^{-N\epsilon^2}$, where $N$ is the number of particles in a fluid parcel. Indeed, in a typical experiment involving water undergoing laminar flow, one may work with fluid parcels of $1$ mm radius, which contain $N \sim 10^{20}$ water molecules. On the other hand, as one pushes relativistic nucleus-nucleus collision simulations to smaller and smaller systems, $N$ progressively decreases, until we reach systems with $N \sim 5$ particle degrees of freedom in a fluid parcel (in pp collisions), where fluctuating phenomena cannot be neglected \cite{Yan:2015lfa}. For this reason, there is increasing interest in including random noise sources in hydrodynamic simulations of the quark-gluon plasma \cite{Singh:2018dpk,Singh:2021mks}. This has also triggered a wave of interesting advancements on the theoretical side \cite{Akamatsu:2016llw,Martinez:2018wia,Nahrgang:2018afz,An:2019osr,An:2020vri,An:2022jgc,Mullins:2023tjg,Mullins:2023bur,Jain:2023obu}.

The present article focuses on the problem of self-consistently adding hydrodynamic fluctuations to the first-order causal and stable theory of relativistic viscous hydrodynamics, known as  BDNK theory (acronym for Bemfica-Disconzi-Noronha-Kovtun \cite{Bemfica2017TheFirst,Bemfica2019_conformal1,Kovtun2019,BemficaDNDefinitivo2020}). This turns out to be an unexpectedly difficult problem, as the standard approaches \cite{Abbasi:2022rum,Mullins:2023ott} suffer from unphysical UV divergences, resulting from a spontaneous ``condensation'' of the non-hydrodynamic modes \cite{GavassinoLyapunov_2020}. Here, we rely on the Fox-Uhlenbeck approach \cite{FoxUhlFluctuations1970} to heal such pathologies, and suppress non-hydrodynamic contributions\footnote{This is similar to the approach employed in \cite{Mullins:2023tjg, Mullins:2023ott} to study stochastic fluctuations in Israel-Stewart theory \cite{Israel_Stewart_1979}. However, we utilize the regularization scheme first presented in \cite{Gavassino:2024ufs} to ensure that the equal-time correlators are well-defined.}. Our main tool to achieve this goal is the following rigorous result (which is proven in section \ref{theBestoBoss}): \textit{If the entropy $S$ entering the microcanonical probability distribution $e^S$ is a Lyapunov functional of the non-fluctuating theory in all reference frames, then the linearised fluctuating theory is causal, stable, and mathematically consistent}. We recall that ``Lyapunov functional'' means ``non-decreasing in time and maximized at equilibrium'' \cite{lasalle1961stability}, and that $S$ is a Lyapunov functional in all reference frames if and only if there is a timelike future-directed vector field $E^\mu$ (known as the information current) such that $\partial_\mu E^\mu \leq 0$, and $S-S_{\text{eq}}=-E={-}\int E^\mu d\Sigma_\mu$ across all Cauchy surfaces \cite{GavassinoGibbs2021,GavassinoCausality2021,GavassinoGENERIC2022,GavassinoStabilityCarter2022}.

Throughout the article we adopt the metric signature $(-,+,+,+)$, and work in natural units: $c=k_B=1$.

\section{Basics of stochastic hydrodynamics}

In this section, we show how one can use the knowledge of the information current to construct a theory of hydrodynamic fluctuations in relativity. In order to do this, we need to ``rederive'' stochastic hydrodynamics, following a path that is slightly different from the textbook derivation \cite{landau5,landau9}, but is equally rigorous, and fully self-contained. 

\subsection{The building blocks of a stochastic theory}

Let $\Psi(x^\mu)\in \mathbb{R}^M$ be a collection of stochastic degrees of freedom of the linearised macroscopic theory, defined so that $\langle\Psi \rangle=0$ at equilibrium (which is assumed to be homogenous). By ``degree of freedom'', here we mean that the knowledge of their value at a given instant of time fully identifies the hydrodynamic state at that time. In the standard approach, the degrees of freedom obey a system of Langevin-type differential equations \cite{landau6}, which are of \textit{first-order} in time:
\begin{equation}\label{Langevin}
    \partial_t \Psi = -\Ma (\partial_j)\Psi +\Sa(\partial_j) \xi \, .
\end{equation}
The field $\xi(x^\mu)\in \mathbb{R}^N$ is a collection of stochastic noises, while the background matrices $\Ma(\partial_j)$ and $\Sa(\partial_j)$ are polynomials in the space derivatives $\partial_j$. In general, the number $M$ of components of $\Psi$ may differ from the number $N$ of components of $\xi$. Hence, while $\Ma$ is a square $M {\times} M$ matrix, $\Sa$ is in general rectangular, with $M {\times} N$ elements.

As part of the definition of $\Psi$ and $\xi$, we assume that $\langle \Psi \rangle$ is the set of ``classical'' (i.e. non-stochastic) hydrodynamic variables, and $\langle \xi \rangle =0$. Hence, when we average \eqref{Langevin}, we recover the classical hydrodynamic equation of motion:
\begin{equation}\label{classical}
   \partial_t \langle \Psi \rangle = -\Ma (\partial_j)\langle \Psi \rangle  \, .
\end{equation}
This implies that, if we are given a classical hydrodynamic theory, then we know the operator $\Ma(\partial_j)$ that enters the stochastic differential equation \eqref{Langevin} of the fluctuating theory. Note that this is true only in the linear regime. In fact, if $\Ma$ depended also $\Psi$ itself, then $\langle\Ma (\partial_j)\Psi\rangle \neq \langle\Ma (\partial_j)\rangle \langle\Psi\rangle$, meaning that non-linear fluctuations ``renormalise'' $\Ma$ \cite{KovtunStickiness2011}.

Let us now focus on the stochastic term $\Sa \xi$ in equation \eqref{Langevin}. There is no general prescription for how many components $\xi$ should have and what the operator $\Sa(\partial_j)$ should look like. Hence, we cannot deduce the whole structure of the noise from hydrodynamics alone, and we may need (in principle) to rely on a microscopic model. However, in practice, we are never interested in the noise itself. Rather, the final goal is to compute hydrodynamic correlators of the form $\langle \Psi \Psi ... \Psi\rangle$. As long as this is our only goal, we don't really need to know the structure of $\Sa$ and $\xi$ in detail. Instead, if the noise is Markovian and the system is in thermodynamic equilibrium, we will be able to use the FDT to effectively get rid of the noise, and obtain formulas for the correlators $\langle \Psi \Psi ... \Psi\rangle$ which are independent of the particular model of noise one is using. The details are provided in the next subsection.


\subsection{The Fox-Uhlenbeck formulation of the fluctuation-dissipation theorem}\label{FDTsection}

Since the equilibrium state is homogenous and stationary, all correlators are invariant under spacetime translations, e.g. $\langle \Psi(x) \Psi^T(y)\rangle=\langle \Psi(x-y) \Psi^T(0)\rangle$. Hence, we can write the two-point correlators of fields and noise as integrals in momentum space, with $k^\mu=(\omega,k^j)$, as follows:
\begin{equation}\label{fourier}
    \begin{split}
\langle \Psi(x) \Psi^T(y)\rangle ={}& \int \dfrac{d^4 k}{(2\pi)^4} e^{i k (x-y)} \Ga^S (k) \, , \\
\langle \xi(x) \xi^T(y)\rangle ={}&  \int \dfrac{d^4 k}{(2\pi)^4} e^{i k (x-y)} \Qa (k) \, , \\
    \end{split}
\end{equation}
where $\Ga^S$ is a Hermitian $M \times M$ matrix, and $\Qa$ is a Hermitian $N \times N$ matrix (recall that $\Psi$ and $\xi$ are column vectors).
If we express \eqref{Langevin} in momentum space, it becomes $\Psi(k)=\big[\Ma(ik_j)-i\omega\big]^{-1} \Sa(ik_j) \xi(k)$. Then, it is straightforward to verify that the correlation matrices $\Ga^S(k)$ and $\Qa(k)$ are related by the following identity \cite{FoxUhlFluctuations1970}:
\begin{equation}\label{fueruzziz}
    \Ga^S(k)=\big[\Ma(ik_j){-}i\omega\big]^{-1} \Sa(ik_j) \Qa(k) \Sa^\dagger(ik_j)\big[\Ma(ik_j){-}i\omega\big]^{-\dagger} \, .
\end{equation}
Note that $\Ma(ik_j){-}i\omega$ is invertible for all $k^\mu \in \mathbb{R}^4/\{0\}$, since the roots of $\det (\Ma-i\omega)$ are the dispersion relations $\omega(k_j)$ of the classical theory, which have a negative imaginary part due to dissipation.
If we plug \eqref{fueruzziz} into the first equation of \eqref{fourier}, and we evaluate the result on two equal-time events, e.g. $x=(0,\x)$ and $y=(0,\y)$, we obtain
\begin{equation}\label{long}
 \langle \Psi(\x) \Psi^T(\y)\rangle = \int \dfrac{d^3 k}{(2\pi)^3} e^{i \kb \cdot (\x-\y)} \int \dfrac{d\omega}{2\pi} \big[\Ma(ik_j){-}i\omega\big]^{-1} \Sa(ik_j) \Qa(k) \Sa^\dagger(ik_j)\big[\Ma(ik_j){-}i\omega\big]^{-\dagger}  \, .   
\end{equation}

Let us now turn our attention to the entropy $S$, which is functional of state, i.e. $S=S[\Psi]$. Since we are working in the linear regime, the fluctuations can be approximated as Gaussian. In other words, we can write $S[\Psi]=S[0]-E[\Psi]$, where $E=\int E^0 d^3 x=\mathcal{O}(\Psi^2)$ is the integral of the zeroth component of the information current \cite{GavassinoCausality2021}, which is quadratic in the fields. Integrating by parts, we can always express $E[\Psi]$ as follows:
\begin{equation}\label{Ewow}
    E = \dfrac{1}{2} \int d^3 x \, \Psi^T \Ka(\partial_j) \Psi \, ,
\end{equation}
where $\Ka(\partial_j)$ is a non-negative definite Hermitian operator. Then, since the probability distribution is $e^S \propto e^{-E}$, all equal-time correlators are Gaussian functional integrals, which can be evaluated using standard field-theory techniques \cite{weinbergQFT_1995}. In the case of two-point correlators, we have
\begin{equation}\label{short}
  \langle \Psi(\x) \Psi^T(\y)\rangle = \dfrac{\int \mathcal{D}\Psi \, e^{-E} \, \Psi(\x) \Psi^T(\y)}{\int \mathcal{D}\Psi \, e^{-E}} = \int \dfrac{d^3 k}{(2\pi)^3} e^{i \kb \cdot (\x-\y)} \Ka^{-1}(ik_j) \, . 
\end{equation}
If we compare \eqref{long} with \eqref{short}, we see that, in order for the noise to be consistent with the equilibrium probability distribution dictated by statistical mechanics, a compatibility constraint must hold:
\begin{equation}\label{eiught}
    \int \dfrac{d\omega}{2\pi} \big[\Ma(ik_j){-}i\omega\big]^{-1} \Sa(ik_j) \Qa(k) \Sa^\dagger(ik_j)\big[\Ma(ik_j){-}i\omega\big]^{-\dagger} = \Ka^{-1}(ik_j) \, .
\end{equation}
Till this point, our analysis was fully general, since no assumption has been made about the noise. Now we introduce the Markovianity postulate, according to which the noise is not correlated in time, i.e. $\langle \xi(x) \xi^T(y)\rangle \propto \delta(x^0-y^0)$. This is equivalent to requiring that $\Qa$ does not depend on $\omega$. Under this assumption, the $\omega$-integral above can be evaluated analytically. With some algebra (which is provided in Appendix \ref{AAA}), one finally arrives at the following identity \cite{FoxUhlFluctuations1970}:
\begin{equation}\label{nuine}
    \Sa(ik_j) \Qa(k_j) \Sa^\dagger(ik_j) = \Ma(ik_j)\Ka^{-1}(ik_j)+\Ka^{-1}(ik_j)\Ma^{\dagger}(ik_j) \, .
\end{equation}
This is the fluctuation-dissipation theorem. It tells us that, even if we do not know the details of $\Sa$ and $\xi$, still, the combination $\Sa\Qa\Sa^\dagger$ is uniquely fixed by statistical mechanics. Plugging \eqref{nuine} into \eqref{fueruzziz}, and \eqref{fueruzziz} into \eqref{fourier}, we finally obtain
\begin{equation}\label{twoPoint}
    \langle \Psi(x) \Psi^T(y)\rangle = \int \dfrac{d^4 k}{(2\pi)^4} e^{i k (x-y)} \big[\Ma(ik_j){-}i\omega\big]^{-1}\big[ \Ma(ik_j)\Ka^{-1}(ik_j){+}\Ka^{-1}(ik_j)\Ma^{\dagger}(ik_j)\big] \big[\Ma(ik_j){-}i\omega\big]^{-\dagger}  \, .
\end{equation}
This shows that (if the noise is Markovian) the two-point correlator is uniquely determined by hydrodynamics alone. In fact, in order to evaluate \eqref{twoPoint}, we only need to know the classical equation of motion \eqref{classical}, which gives us $\Ma$, and the information current, which gives us $\Ka$ through equation \eqref{Ewow}.

\subsection{The forward and the backward correlator}

It is straightforward to verify that the symmetric correlator $\Ga^S(x-y)=\langle \Psi(x)\Psi^T(y)\rangle$, as given in \eqref{twoPoint}, is the sum of two correlators: the forward correlator $\Ga^+(x{-}y)$, and and the backward correlator $\Ga^-(x{-}y)$, given by, respectively,
\begin{equation}\label{cacioppoli}
\begin{split}
\Ga^+(x-y)={}& \int \dfrac{d^4 k}{(2\pi)^4} e^{i k (x-y)} \big[\Ma(ik_j)-i\omega \big]^{-1}\Ka^{-1}(ik_j) \, , \\
\Ga^-(x-y)={}& \int \dfrac{d^4 k}{(2\pi)^4} e^{i k (x-y)} \Ka^{-1}(ik_j) \big[\Ma(ik_j)-i\omega \big]^{-\dagger} \, . \\
\end{split}
\end{equation}
Recalling that $\Ka(ik_j)$ is Hermitian, we see that $\Ga^-(x-y)=\Ga^+(y-x)^T$. Furthermore, using the first equation in \eqref{freesbone}, we can perform the integral in $\omega$ analytically, and we obtain:
\begin{equation}\label{retuzinfinie}
   \Ga^+(x-y)  = \Theta(x^0{-}y^0) \int \dfrac{d^3 k}{(2\pi)^3} e^{i\textbf{k}\cdot(\textbf{x}-\textbf{y})} \, e^{-(x^0-y^0)\Ma(ik_j)}\Ka(ik_j)^{-1} \, .   
\end{equation}
This tells us that $\Ga^+$ and $\Ga^-$ are just the restrictions of $\Ga^S$ to positive and negative times respectively, namely $\Ga^\pm(x)=\Theta(\pm \, x^0)\langle \Psi(x)\Psi^T(0)\rangle$. Furthermore, we note that, for $x^0>0$, $\Ga^+(x)$ is a solution of the classical equation of motion \eqref{classical}, with initial condition \eqref{short} (see appendix \ref{freestreaming} for a more direct proof) \cite{landau5}. Thus, the forward correlator is a solution of the following partial differential equation:
\begin{equation}\label{fubini}
    \big[\partial_t +\Ma(\partial_j)\big] \Ga^+(x) = \delta(t) \Ga^S(\textbf{x}) \, .
\end{equation}

\subsection{The Martin-Siggia-Rose approach}

Even if we know the two-point correlator, we still need a procedure to evaluate all higher-point correlators $\langle \Psi \Psi...\Psi\rangle$. Such a procedure is called Martin-Siggia-Rose (MSR) approach \cite{Martin:1973zz,Granese:2022igc}, and is summarised below.

Our goal is to find the probability distribution $\mathcal{P}[\Psi]$ for the ``history'' of the field $\Psi(x^\mu)$ in a spacetime region of interest. To this end, let us first note that $\Psi$ is a solution of the stochastic differential equation \eqref{Langevin}.
Hence, if we fix an initial condition $\Psi_{-\infty}$ in the far past, we can think of $\Psi$ as a functional $\Psi[\Psi_{-\infty},\xi]$, in the sense that for every choice of $\Psi_{-\infty}$ and $\xi(x^\mu)$ there is a unique realization of the field $\Psi(x^\mu)$ that solves \eqref{Langevin}\footnote{The existence of the information current implies that the solutions to \eqref{Langevin} are unique. Proof: If $\Psi_1$ and $\Psi_2$ are two solutions of \eqref{Langevin} with the same initial condition and the same $\xi$, then $\Delta \Psi=\Psi_2{-}\Psi_1$ is a solution of \eqref{Langevin} with vanishing initial condition and $\xi=0$. Thus, $E[\Delta \Psi]{=}0$ in the far past. Since $E$ is non-negative definite, and it is non-increasing along solutions of \eqref{Langevin} without noise, we have $E[\Delta \Psi]{=}0$ at all later times. But $E[\Delta \Psi]$ vanishes if and only if $\Delta \Psi{=}0$ (by definition of Lyapunov function \cite{lasalle1961stability}), proving that $\Psi_2=\Psi_1$.}. Thus, we can write
\begin{equation}
    \mathcal{P}[\Psi]= \int \mathcal{D} \Psi_{-\infty} \, \mathcal{P}[\Psi_{-\infty}] \int \mathcal{D}\xi \,   \mathcal{P}[\xi] \, \delta^{\infty}\big[\Psi-\Psi[\Psi_{-\infty},\xi]\big] \, ,
\end{equation}
where $\delta^{\infty}$ is a functional Dirac delta. However, since the system is dissipative, it quickly loses memory of its initial state in the far past, so that we can write $\Psi[\Psi_{-\infty},\xi]=\Psi[\xi]$ (in the spacetime region of interest). Therefore, we can directly perform the integral over the initial data, and we are left with
\begin{equation}\label{veciopreston}
    \mathcal{P}[\Psi]=  \int \mathcal{D}\xi \,   \mathcal{P}[\xi] \, \delta^{\infty}\big[\Psi-\Psi[\xi]\big] \, ,
\end{equation}
Considering that $\Psi=\Psi[\xi]$ if and only if \eqref{Langevin} holds, we know that $\delta^{\infty}\big[\Psi-\Psi[\xi]\big]$ and $\delta^{\infty}\big[\partial_t \Psi {+}\Ma\Psi {-}\Sa\xi\big]$ are proportional to each other. The proportionality constant is the functional determinant of $\partial_t{+}\Ma$  (see appendix \ref{BBB}), which does not depend on $\Psi$ and $\xi$. It is therefore a global constant, which can be taken out of all integrals. Thus, we have
\begin{equation}
    \mathcal{P}[\Psi] \propto \int \mathcal{D}\xi \,   \mathcal{P}[\xi] \int \mathcal{D} \Tilde{\Psi} \, \, e^{i\int d^4x \Tilde{\Psi}^T(\partial_t \Psi {+}\Ma\Psi {-}\Sa\xi)} \, , 
\end{equation}
where we employed the identity $2\pi\delta(a)=\int e^{ia\tilde{x}} d\tilde{x}$ to convert the Dirac delta into a Fourier integral, at the expense of introducing a new fictitious field $\Tilde{\Psi}(x^\mu)$. Now, even if we do not know the probability distribution $\mathcal{P}[\xi]$, we know that it is Gaussian, because we are working in the linear regime. Hence, if we perform the integral in the noise, the most general result must have the form of a bilocal Gaussian weight for the variable $\Tilde{\Psi}$ (to see this, just complete the square). Thus, we arrive at the generic expression
\begin{equation}\label{probuzzuz}
    \mathcal{P}[\Psi] \propto \int \mathcal{D} \Tilde{\Psi} \, \, e^{i\int d^4x \Tilde{\Psi}^T(\partial_t \Psi {+}\Ma\Psi) -\frac{1}{2}\int d^4x \, d^4y \, \Tilde{\Psi}^T(x) \Ha(x-y)\Tilde{\Psi}(y)} \, ,
\end{equation}
where $\Ha$ is some bilocal kernel which depends only on the difference $x-y$ due to translation invariance. This implies that the average of an arbitrary functional $F[\Psi]$ can be expressed as a path integral as follows:
\begin{equation}\label{pathintegral}
    \langle F[\Psi] \rangle = \dfrac{\int \mathcal{D}\Psi \mathcal{D}\Tilde{\Psi} \, e^{i \mathcal{S}} \, F[\Psi]}{\int \mathcal{D}\Psi \mathcal{D}\Tilde{\Psi} \, e^{i\mathcal{S}} } \, ,
\end{equation}
with the (complex) effective action
\begin{equation}\label{effectiveaction}
    \mathcal{S}[\Psi,\Tilde{\Psi}]=\int d^4 x \, \Tilde{\Psi}^T(\partial_t \Psi {+}\Ma\Psi) +\frac{i}{2}\int d^4x \, d^4y \, \Tilde{\Psi}^T(x) \Ha(x{-}y)\Tilde{\Psi}(y) \, .
\end{equation}
To complete the construction of the theory, we only need to specify the kernel $\Ha(x-y)$. This is easily done by requiring that the two-point correlators, as computed through the path integral \eqref{pathintegral}, coincide with the explicit formula \eqref{twoPoint} provided by the fluctuation-dissipation theorem. All the detailed calculations are reported in appendix \ref{BBB2}. Here, we provide only the final result:
\begin{equation}\label{fbruf}
    \Ha(x{-}y)=\delta(x^0{-}y^0) \int \dfrac{d^3 k}{(2\pi)^3} e^{i \kb \cdot (\x-\y)} \big[ \Ma(ik_j)\Ka^{-1}(ik_j)+\Ka^{-1}(ik_j)\Ma^{\dagger}(ik_j)\big] \, .
\end{equation}
Again, we see that the knowledge of the classical equation of motion and of the information current fully specifies the stochastic theory. In fact, if we fix the operators $\Ma(\partial_j)$ and $\Ka(\partial_j)$, this uniquely identifies the effective action $\mathcal{S}$, thereby allowing us to compute all correlators $\langle \Psi \Psi ... \Psi\rangle$. 

Let us remark that the presence of the factor $\delta(x^0{-}y^0)$ in equation \eqref{fbruf} is related to the Markovianity assumption for the noise. If we released this assumption, $\Ha$ would depend on $x^0{-}y^0$ in non-trivial ways.

\section{Mathematical consistency, causality, and stability}\label{theBestoBoss}

It is well-known that, if the classical equation of motion \eqref{classical} admits an information current, then it is causal and covariantly stable \cite{GavassinoGibbs2021,GavassinoCausality2021}. Furthermore, it is straightforward to prove that the initial value problem is uniquely solvable for initial data in the Schwartz space \cite{Hishcock1983,GavassinoUniversality2023}. Hence, the classical theory is well-behaved and mathematically consistent. In this section, we show that similar results hold also for the fluctuating theory. 

\subsection{Convergence of the functional integrals}

One of the major difficulties with fluctuating hydrodynamics is that, even if a hydrodynamic theory is stable on-shell, namely along solutions of \eqref{classical}, it may become unstable off-shell, if certain (usually UV) fluctuations are entropically favored \cite{GavassinoCausality2021}. When this happens, the probability distribution $\mathcal{P}[\Psi]$, as given in \eqref{probuzzuz}, is non-normalizable \cite{Mullins:2023ott}. As a consequence, the MSR path integral \eqref{pathintegral} becomes ill-defined, and the whole fluctuating theory is mathematically inconsistent. Let us show that, in the presence of a well-defined information current, this can never happen.

First, let us note that, if the information current is timelike future directed, then $E$ is non-negative definite. Hence, $e^{-E}$ is a regular Gaussian distribution, which is normalizable. Thus, the functional integral \eqref{short} is well-defined, and all equal-time correlators are ``convergent'' (in a distributional sense). The non-equal-time ones require more work.

Let $\Phi(x^\mu)$ be a solution of the classical equation of motion \eqref{classical}. Then, along $\Phi$, we have that $\partial_\mu E^\mu \leq 0$. Integrating over the volume, and switching to momentum space, we obtain
\begin{equation}\label{lkjhgfds}
    \dfrac{d}{dt} \int \dfrac{d^3 k}{(2\pi)^3} \, \Phi^\dagger(t,\textbf{k}) \Ka(i\textbf{k})\Phi(t,\textbf{k}) \leq 0 \, .
\end{equation}
Expressing the initial condition for $\Phi(t,\textbf{k})$ in the form $\Phi(0,\textbf{k})=\Ka(i\textbf{k})^{-1}V(\textbf{k})$, we can solve \eqref{classical} explicitly, and get $\Phi(t)= e^{-\Ma t}\Ka^{-1}V$, where the $\textbf{k}-$dependence is from now on understood. Then, \eqref{lkjhgfds} becomes
\begin{equation}
 \dfrac{d}{dt} \int \dfrac{d^3 k}{(2\pi)^3} \, V^\dagger \, \Ka^{-1} \, e^{-\Ma^\dagger t} \, \Ka \, e^{-\Ma t} \, \Ka^{-1} V \leq 0 \, .   
\end{equation}
Evaluating this formula at $t=0$, we finally obtain
\begin{equation}\label{goofy}
  0 \leq  \int \dfrac{d^3 k}{(2\pi)^3} V^\dagger (\Ma \Ka^{-1}{+}\Ka^{-1}\Ma^\dagger) V = \int d^3 x \, d^4 y \, V^T(x) \Ha(x-y) V(y)   \, ,
\end{equation}
where we used equation \eqref{fbruf} in the last step. But since $V(\textbf{x})=\Ka(\partial_j)\Phi(\textbf{x})$ is arbitrary (as $\Ka$ is invertible), we can conclude that $\mathfrak{Im} \mathcal{S}$ is always non-negative, see equation \eqref{effectiveaction}. It follows that the weight $|e^{i\mathcal{S}}|=e^{\mathfrak{-Im} \mathcal{S}}$ is a regular Gaussian, making the probability distribution \eqref{probuzzuz} normalizable, and the path integral \eqref{pathintegral} well-defined.

As a consistency check, we also note that, due to the first inequality in \eqref{goofy}, and identity \eqref{twoPoint}, the Hermitian matrix $\Ga^S(k)$ is necessarily non-negative definite, as it should be, since $\langle \Psi(k)\Psi^\dagger(p)\rangle=(2\pi)^4 \delta^4(k-p) \Ga^S(k)$. Previous attempts at deriving a fluctuating generalization of BDNK failed at making $\Ga^S(k)$ non-negative definite \cite{Abbasi:2022rum,Mullins:2023ott}, and are strictly speaking mathematically inconsistent. However, if one has an information current and follows our procedure, the resulting $\Ga^S(k)$ is non-negative definite by construction.

\subsection{Bounds from microcausality}\label{micri}

The dispersion relations $\omega_n(\textbf{k})$ of the fluctuating theory are \textit{defined} to be the singularities of the forward correlator in momentum space. From equation \eqref{retuzinfinie}, we see that the such singularities are determined by the algebraic equation $\det \big[ \Ma(i\textbf{k})-i\omega \big]=0$, which is also the defining equation for the dispersion relations of the classical theory \eqref{classical}. But if there is an information current, the classical theory is causal and covariantly stable, so that the dispersion relations satisfy all the necessary conditions for covariant stability. In particular, setting $\textbf{k}=(k^x,0,0)$, we have \cite{GavassinoBounds2023}
\begin{equation}
    \mathfrak{Im} \, \omega_n \leq |\mathfrak{Im} \, k^x| \, ,
\end{equation}
which is also the consistency criterion with microcausality in quantum field theory. Hence, the dispersion relations of the fluctuating theory obey all the causality bounds derived in \cite{HellerBounds2022}.

\subsection{Causality and covariant stability of the fluctuating theory}\label{macrY}

Now we only need to prove that the full fluctuating theory is causal and stable. To this end, let us couple the system with a small external (non-stochastic) forcing term $F(x^\mu)$, which drives the fluid out of equilibrium. Then, the Langevin-type equation of motion \eqref{Langevin} becomes
\begin{equation}\label{langevi2}
     \partial_t \Psi = -\Ma (\partial_j)\Psi +\Sa(\partial_j) \xi +F \, ,
\end{equation}
whose solution is
\begin{equation}\label{realization}
    \Psi(x)= \int \dfrac{d^3 k}{(2\pi)^3} e^{i\textbf{k}\cdot\textbf{x}} \! \! \! \int_{-\infty}^t \! \! \! \! ds \,  e^{-\Ma (t-s)} \Sa \xi(s) \; + \; \int \dfrac{d^3 k}{(2\pi)^3} e^{i\textbf{k} \cdot\textbf{x}}  \! \! \! \int_{-\infty}^t \! \! \! \! ds \,  e^{-\Ma (t-s)} F(s) \, ,
\end{equation}
where, again, the dependence of the integrand on $\textbf{k}$ is understood. As can be seen, the solution to \eqref{langevi2} is the sum of two contributions. The first, which we call $\Psi_u$, is just the unperturbed solution of \eqref{Langevin}, i.e. the solution in the absence of external forces. The second, which we call $\Psi_p$, is the \textit{classical} perturbation introduced by $F$, and is a solution of the non-stochastic equation of motion $\partial_t \Psi_p =-\Ma \Psi_p +F$.

In the stochastic theory, an individual realization \eqref{realization} of the system does not teach us anything about causality and stability. In fact, apparent superluminal signals (or apparent unstable bursts) may just be the result of random fluctuations \cite{DoreGavassino2022}. Instead, to assess these features rigorously, one has to repeat the same experiment multiple times, using the same $F$, and study the impact of the external forcing on the correlators $\langle \Psi \Psi...\Psi\rangle$. From \eqref{realization}, we have
\begin{figure}
\begin{center}
\includegraphics[width=0.55\textwidth]{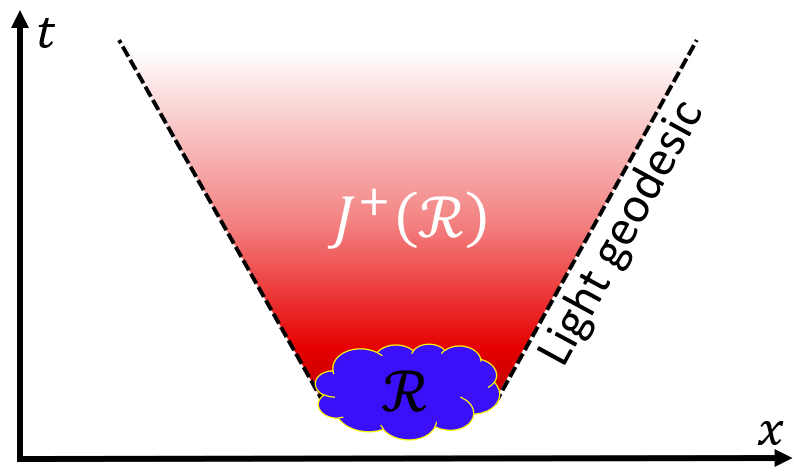}
	\caption{Qualitative Minkowski diagram of the perturbation $\Psi_p(x^\mu)$, which is a classical solution of the non-stochastic differential equation $\partial_t \Psi_p = -\Ma(\partial_j)\Psi_p+F$, where $F(x^\mu)$ is an external forcing term. The blue region $\mathcal{R}$ is the (compact) support of $F$, while the shades of red are a colormap of the intensity $||\Psi_p||^2$ (red=large, white=small). From causality of the non-stochastic equation of motion \eqref{classical}, we know that the support of $\Psi_p$ is contained inside $J^+(\mathcal{R})$, namely the future lightcone of $\mathcal{R}$ \cite{Susskind1969,Wald,Hawking1973}. From stability+dissipation of \eqref{classical}, we also know that $\Psi_p \rightarrow 0$ at late times \cite{GavassinoSuperluminal2021}.}
	\label{fig:JR}
	\end{center}
\end{figure}
\begin{equation}\label{piusemprepiu}
    \langle \Psi \Psi ... \Psi \rangle = \langle (\Psi_u{+}\Psi_p)(\Psi_u{+}\Psi_p)...(\Psi_u{+}\Psi_p)\rangle \, .
\end{equation}
Now, let us assume that $F(x^\mu)$ is supported inside a compact spacetime region $\mathcal{R}$. Then, since the classical theory \eqref{classical} is causal and covariantly stable (because we have an information current \cite{GavassinoGibbs2021,GavassinoCausality2021}), and there is dissipation, the Minkowski diagram of $\Psi_p(x^\mu)$ must possess certain mathematical properties \cite{GavassinoSuperluminal2021}, which are schematically illustrated in figure \ref{fig:JR}. First, $\Psi_p$ is supported inside $J^+(\mathcal{R})$, so that, if we consider correlators that are causally disconnected from $\mathcal{R}$, the right-hand side of \eqref{piusemprepiu} reduces to $\langle \Psi_u \Psi_u ... \Psi_u \rangle $. This tells us that the information about $F$ cannot exit the lightcone, meaning that the stochastic theory is rigorously causal. Secondly, $\Psi_p$ decays to zero at late times, so that $\langle \Psi \Psi ... \Psi \rangle \rightarrow \langle \Psi_u \Psi_u ... \Psi_u \rangle$ long time after the force $F$ has acted. This means that the equilibrium state is stable in the rest frame. But since all Lorentz observes agree on whether a subluminal signal grows or decays, stability assessments are Lorentz-invariant in causal systems. In particular, following the same steps as in the proof of Theorem 2 of \cite{GavassinoSuperluminal2021}, one can straightforwardly establish \textit{covariant} stability of the full stochastic theory. 

\subsection{More on the Fluctuation-Dissipation Theorem}

In section \ref{FDTsection}, we discussed the Fox-Uhlembeck formulation of the FDT in relativistic hydrodynamics. However, the most general formulation of the fluctuation-dissipation theorem (in the Gaussian limit) is an identity relating the symmetric correlator $\Ga^S$ to the retarded linear response Green function of the system. Let us show that the two formulations of the FDT are indeed equivalent.

Let us introduce the auxiliary field $\Lambda=\Ka(\partial_j)\Psi$, which is the thermodynamical conjugate of $\Psi$, see equation \eqref{Ewow}. If we Fourier-transform in space, we can rewrite the classical equation of motion \eqref{classical} in Onsager's canonical form $\partial_t \langle \Psi\rangle =-\Ma(ik_j)\Ka^{-1}(ik_j)\langle \Lambda \rangle $. Following \citet{landau5}, if we couple $\Psi$ with an external force $f$ through an interaction potential of the form $-\Psi^T f$, then the externally perturbed equation of motion can be obtained through the replacement $\langle \Lambda \rangle \rightarrow \langle \Lambda \rangle -f/T$, where $T$ is the temperature of the background state. The result is
\begin{equation}\label{shinzkone}
    \partial_t \langle \Psi \rangle = -\Ma \Ka^{-1} \big(\langle\Lambda\rangle -f/T \big) \, .
\end{equation}
This equation is the statistical average of \eqref{langevi2}, once one makes the identification $F=\Ma \Ka^{-1} f/T$.  Fourier-transforming \eqref{shinzkone} also in time, we obtain an equation of the form $\langle \Psi(k)\rangle = \Ga^R(k) f(k)$, with retarded Green's functions\footnote{The poles of the retarded Green's function are the same as those of the forward correlator obtained from equation \eqref{cacioppoli}. If $\Psi$ is a list of local observables, then the retarded correlator $\Ga^R(x)$ in real space must vanish outside the lightcone by causality \cite{HellerBounds2022}. It turns out that, in BDNK, this is usually not the case, and $\Ga^R(x)$ is non-vanishing for $t>0$ everywhere. This implies that some (but not all) components of $\Psi$ are non-local. Such components correspond to quantities that depend on the choice of hydrodynamic frame (and have no physical significance \cite{Kovtun2019}). See section \ref{reparam} for more details.} 
\begin{equation}
    \Ga^R(k)=(\Ma-i\omega)^{-1} \dfrac{\Ma \Ka^{-1}}{T} \, .
\end{equation}
Comparing with equation \eqref{twoPoint}, we recover the FDT in its standard formulation \cite{kovtun_lectures_2012}:
\begin{equation}
    \Ga^S(k)=\dfrac{T}{i\omega} \bigg[\Ga^R(k)-\Ga^R(k)^\dagger \bigg] \, . 
\end{equation}

\section{Relativistic diffusion}\label{nondiffidate}

Now that the consistency of the theory has been established, we can provide some concrete examples. Our first example is the relativistic diffusion equation, interpreted as a first-order (BDNK-type) theory. Its interpretation as an Israel-Stewart theory, and the related fluctuating generalization, have already been studied in detail \cite{Mullins:2023tjg,Mullins:2023ott}.

\subsection{Classical theory}

Let $\varphi(x^\mu) \in \mathbb{R}$ be a linearised stochastic field, with units $L^{-3/2}$ ($L$ being ``length''). Suppose that its classical equation of motion is $\tau \partial^2_t\langle \varphi \rangle +\partial_t \langle \varphi \rangle =D \partial_j \partial^j \langle \varphi \rangle$, where $D$ is a diffusion constant and $\tau$ is the UV cutoff timescale of the theory.  This equation of motion is of second order in time. It follows that the system has two dynamical degrees of freedom. Hence, we need to introduce a second stochastic variable $\uppi(x^\mu)$, such that
\begin{equation}\label{classuzzuz}
    \begin{split}
& \tau \partial_t \langle \varphi \rangle +\langle \varphi \rangle = \langle \uppi \rangle \, , \\
& \partial_t \langle \uppi \rangle = D \partial_j \partial^j \langle \varphi \rangle \, . \\
    \end{split}
\end{equation}
From this, we immediately obtain the evolution operator for the couple $\Psi =\{\varphi,\uppi\}^T$, namely
\begin{equation}
    \Ma(\partial_j) = 
    \begin{bmatrix}
        1/\tau & -1/\tau \\
        -D \partial_j \partial^j &  0 \\
    \end{bmatrix} \, .
\end{equation}
Now we need the information current. In \cite{Gavassino:2024ufs}, it was shown that the UV-regularised information current for this theory is not unique. However, different regularization prescriptions have different regimes of applicability, with different cutoff lengthscales $\lambda$. The one with shortest cutoff lengthscale is the one with $\lambda=\tau$, which, expressed in terms of $\varphi$ and $\uppi$, reads explicitly
\begin{equation}\label{INFOONA}
\begin{split}
E^0 ={}& \dfrac{1}{2} \big[ \uppi^2+ \tau  D \partial_j \varphi \partial^j \varphi \big] \, ,\\
E^j ={}& -D \, \uppi \, \partial^j \varphi \, ,\\
\sigma ={}&  D \partial_j \varphi \partial^j \varphi \, , \\
\end{split}
\end{equation}
where $\sigma$ is the associated entropy production rate. Note that, since $\varphi$ and $\uppi$ have dimensions $L^{-3/2}$, $E^\mu$ has dimension $L^{-3}$, as it is the density of a dimensionless quantity (the information). It can be easily checked that the dissipation equation $\partial_\mu E^\mu=-\sigma$ holds along all solutions of the classical equation of motion \eqref{classuzzuz}. The conditions for $E^\mu$ to be timelike future-directed, and for $\sigma$ to be positive definite, are $\tau \geq D>0$. Under these conditions, both the classical and the fluctuating theory are mathematically consistent, causal, and covariantly stable. The condition $\tau \geq D$ tells us that the cutoff timescale of the theory cannot be shorter than the diffusion step $D$ \cite{Hartman:2017hhp}. From \eqref{INFOONA}, we obtain
\begin{equation}
    \Ka(\partial_j) = 
    \begin{bmatrix}
        -\tau D \partial_j \partial^j & 0\\
        0 & 1 \\
    \end{bmatrix} \, .
\end{equation}

\subsection{Fluctuating theory}\label{flucttuuttuioz}

From the knowledge of $\Ma$ and $\Ka$, we can build the whole fluctuating theory. From \eqref{short}, we can straightforwardly compute the two-point equal-time correlators. Solving the integral in momentum directly, we obtain
\begin{equation}\label{grzuzber}
    \begin{split}
 \langle \varphi(\x)\varphi(\y)\rangle ={}& \dfrac{1}{4\pi \tau D |\x -\y|}  \, , \\
 \langle \varphi(\x)\uppi(\y)\rangle ={}&  0 \, , \\
 \langle \uppi(\x)\uppi(\y)\rangle ={}& \delta^3(\x-\y)  \, . \\
    \end{split}
\end{equation}
From equations \eqref{effectiveaction} and \eqref{fbruf}, we can compute the MSR action. If we express it as the spacetime integral of a Lagrangian density, i.e. $\mathcal{S}=\int \mathcal{L}(x)d^4x$, we find
\begin{equation}\label{lagringa}
    \mathcal{L} = \dfrac{\tilde{\varphi}}{\tau} \big[\varphi+\tau \partial_t \varphi-\uppi\big] +\tilde{\uppi} \big[ \partial_t \uppi-D \partial_j \partial^j \varphi  \big] +\dfrac{i \tilde{\varphi}}{4\pi \tau^2 D}  \int  \dfrac{\tilde{\varphi}(x^0,\y)}{|\x -\y|} d^3 y \, .
\end{equation}
The symmetric two-point correlators at non-equal times can be computed from equation \eqref{twoPoint}:
\begin{equation}\label{nonGyzxiop}
     \begin{bmatrix}
      \langle \varphi(x) \varphi(y)\rangle & \langle \varphi(x) \uppi(y)\rangle  \\
      \langle \uppi(x) \varphi(y)\rangle  & \langle \uppi(x) \uppi(y)\rangle \\
     \end{bmatrix}  = \int \dfrac{d^4 k}{(2\pi)^4} \dfrac{ 2 e^{i k (x-y)} }{\omega^2{+}(D\kb^2 {-}\tau \omega^2)^2} 
     \begin{bmatrix}
        \omega^2(D\kb^2)^{-1} & i\omega \\
        -i\omega  & D\kb^2 \\ 
     \end{bmatrix}\, .
\end{equation}
Finally, the forward correlators can be computed from equation \eqref{cacioppoli}. The result is
\begin{equation}\label{inritardo}
\Theta(x^0{-}y^0)
     \begin{bmatrix} 
      \langle \varphi(x) \varphi(y)\rangle & \langle \varphi(x) \uppi(y)\rangle  \\
      \langle \uppi(x) \varphi(y)\rangle  & \langle \uppi(x) \uppi(y)\rangle \\
     \end{bmatrix}  = \int \dfrac{d^4 k}{(2\pi)^4} \dfrac{e^{i k (x-y)} }{-i\omega+D\kb^2-\tau \omega^2} 
     \begin{bmatrix}
        -i\omega(D\kb^2)^{-1} & 1 \\
        -1  & 1-i\omega \tau \\ 
     \end{bmatrix}\, .
\end{equation}
The matrix exponential in \eqref{retuzinfinie} can be computed explicitly, but the result is rather cumbersome, and not very enlightening. Usually, one is not able to provide an analytic expression for the correlators at non-equal times, which are then left in the form of integrals in momentum space. This is not a problem for practical applications, since one is mostly interested in the infrared (i.e. small $\kb$) part of the Fourier integrals.

\subsection{Interpretation of the classical theory}

We aim to analyse the physical content of the fluctuating theory constructed above. To this end, we need to fix a precise physical interpretation for the variables $\varphi$ and $\uppi$. This is easily done once we note that the classical equations of motion \eqref{classuzzuz} can be rewritten in the form of a conservation law, $\partial_\mu \langle J^\mu \rangle =0$, with conserved current
\begin{equation}\label{napoeloneee}
    J^\mu \propto
    \begin{pmatrix}
        \uppi \\
        -D \partial^j \varphi \\
    \end{pmatrix}
    +\text{ ``noise'' }\, .
\end{equation}
Therefore, $\uppi$ can be interpreted as the fluctuation to a conserved density $n$, and $\varphi$ as the fluctuation to the associated chemical potential $\mu$ (recall Fick's law: $\textbf{J} \propto -\nabla \mu$). Of course, both $\varphi$ and $\uppi$ have been rescaled by a background constant, in order to have units $L^{-3/2}$. In particular, if $T$ is the temperature, we have that
\begin{equation}
    \begin{split}
        \varphi ={}& \bigg(\dfrac{1}{T}\dfrac{\partial n}{d\mu} \bigg)^{1/2} \delta \mu \, , \\
        \uppi ={}&  \bigg(\dfrac{1}{T}\dfrac{d\mu}{dn} \bigg)^{1/2} \delta n \, . \\
    \end{split}
\end{equation}
Now, from the first equation of \eqref{classuzzuz}, we see that $\delta n$ and $\delta \mu$ are not proportional to each other, and are independent stochastic degrees of freedom. Thus, the model under consideration is a BDNK theory of diffusion, whose conserved current has the following (first-order) classical constitutive relation \cite{Kovtun2019}:
\begin{equation}
  \langle J^\mu \rangle  = \dfrac{dn}{d\mu}
    \begin{pmatrix}
        \langle\delta \mu\rangle {+}\tau \partial_t \langle\delta \mu\rangle \\
        -D \partial^j \langle\delta \mu \rangle  \\
    \end{pmatrix}   \, .
\end{equation}
Starting from this interpretation, let us discuss the results of section \ref{flucttuuttuioz}. 

\subsection{Analysis of the fluctuating theory}

Equation \eqref{grzuzber} tells us that (at a given instant of time) the fluctuations in chemical potential are long range, since $\langle \delta \mu(\x) \delta \mu(\y) \rangle \propto |\x-\y|^{-1}$, and they are uncorrelated from the fluctuations in density, since $\langle \delta n \, \delta \mu \rangle=0$. Both these phenomena arise because the UV sector of BDNK contains the spurious non-hydrodynamic mode $\varphi(x^\mu)=e^{-t/\tau}$, which describes a pure ``frame relaxation'' that does not involve any transport of charge (i.e. $\uppi=0$). In the classical theory, this mode is irrelevant, being dissipated away within a timescale $\tau$. However, in the fluctuating theory, it can be repeatedly activated by the noise, causing the chemical potential to perform spontaneous jumps of the kind $\delta \mu \rightarrow \delta \mu +a \, e^{-t/\tau}$ ($a=\text{const}$). Since these jumps are uniform in space, they induce long-range correlations in the chemical potential. Furthermore, since they leave $\delta n$ unaffected, they statistically decouple $\delta \mu$ from $\delta n$. For this reason, the equal-time density correlator remains local, and it coincides with that of the non-relativistic theory,
\begin{equation}
    \langle \delta n(\x) \delta  n (\y)\rangle = T \dfrac{dn}{d\mu} \delta^3(\x-\y) \, .
\end{equation}

Let us now turn our attention to the MSR Lagrangian density \eqref{lagringa}. As can be seen, it is local in time, but non-local in space. The locality in time is a consequence of the Markovianity assumption for the noise. The non-locality in space appears because $\langle \delta \mu(\x) \delta \mu(\y) \rangle$ is long-range. We remark that the non-locality of the effective action is not a signal of causality violation, because ``long-range correlation'' is not equivalent to ``long-range causation'', and the fluctuating theory is causal by construction, see sections \ref{micri} and \ref{macrY}. Also, it is automatically verified that $\mathfrak{Im} \, \mathcal{S} \geq 0$, since we can write $\mathfrak{Im} \, \mathcal{S}=\int U[\tilde{\varphi}] dt$, where
\begin{equation}
    U[\tilde{\varphi}]= \dfrac{1}{4\pi \tau^2 D}\int d^3x \, d^3 y \, \dfrac{\tilde{\varphi}(\x)\tilde{\varphi}(\y)}{|\x-\y|} 
\end{equation}
is analogous to the Coulomb potential energy, which is non-negative definite \cite{jackson_classical_1999}. Thus, we have resolved the problem of mathematical inconsistency of fluctuating BDNK discussed in \cite{Mullins:2023tjg}

Finally, let us focus on the non-equal time correlator \eqref{nonGyzxiop}. Since the chemical potential undergoes unphysical ``hydrodynamic-frame fluctuations'', we cannot learn much from it. Instead, we should only focus on the density-density correlator, which is what can be eventually measured in an experiment:
\begin{equation}\label{surprise!}
    \langle \delta n(x) \delta n(y) \rangle = T \dfrac{dn}{d\mu} \int \dfrac{d^4k}{(2\pi)^4} e^{ik(x-y)} \dfrac{2D \kb^2}{\omega^2+(D\kb^2-\tau \omega^2)^2} \, .
\end{equation}
We note that, in the infrared limit ($\omega\tau \ll 1$), this correlator reduces to its non-relativistic analogue, see \cite{landau9}, \S 89, Problem 1. More importantly, the correlator \eqref{surprise!} coincides with the density-density correlator of the Israel-Stewart theory of diffusion \cite{Mullins:2023tjg}.

\subsection{The noise is not white}

From equation \eqref{nuine}, we find that 
\begin{equation}\label{grufjuion}
   \Sa(i\kb) \Qa(k) \Sa(i\kb)^\dagger= \dfrac{2}{\tau^2 D \kb^2} 
    \begin{bmatrix}
        1 & 0 \\
        0 &  0 \\
    \end{bmatrix} \, ,
\end{equation}
where we recall that $\Qa(k)$ is the Fourier transform of the noise-noise correlator, see equation \eqref{fourier}. But since $\Sa$ is expected to be a polynomial in the wavector, $\Qa(k)$ must contain a power $\kb^{-2}$ or lower, for equation \eqref{grufjuion} to hold. It follows that $\langle \xi(\x)\xi(\y) \rangle$ cannot be a simple Dirac delta in space, and it must be long-range. This tells us that, even if we did not specify a model for the noise, we know that it cannot be a white noise in space.

\section{Non-dispersive telegrapher equation}

In 3+1 dimensions, it is usually impossible to find an analytical expression for the forward correlator $\Ga^+(x)$ in real space. Hence, one needs to keep the Fourier integral \eqref{cacioppoli} unsolved, or to rely on numerical techniques. However, it is useful to have a least one explicit example where the Fourier integral can be solved analytically, which is what we provide in this section.

\subsection{Outline of the theory}

If $\varphi$ is the chemical potential of the electric charge, then the relativistic diffusion equation is coupled to the Maxwell equations, and we must introduce an Ohmic correction to the four-current. The interaction with the electromagnetic field modifies the relativistic diffusion equation into $\tau \partial^2_t \langle \varphi \rangle +(1{+}\tau \Sigma)\partial_t \langle \varphi \rangle+\Sigma \langle \varphi \rangle = D\partial_j \partial^j \langle \varphi \rangle$, where $\Sigma$ is the electric conductivity coefficient \cite{Gavassino:2024ufs}. If we set $\tau=D=\Sigma=1$, we obtain the following equation of motion:
\begin{equation}\label{telluz}
     \partial^2_t \langle \varphi \rangle +2\partial_t \langle \varphi \rangle+ \langle \varphi \rangle = \partial_j \partial^j \langle \varphi \rangle \, .
\end{equation}
The above choice of transport coefficients is particularly convenient, as the dispersion relations are just $\omega{=}-i\pm |\kb|$. Equation \eqref{telluz} models the propagation of electric pulses through non-dispersive dissipative media, and is frequently used in telegraphy \cite{CourantHilbert2_book}. As we did in section \ref{nondiffidate}, we introduce an auxiliary stochastic degree of freedom $\uppi$, such that
\begin{equation}\label{equiboo}
\begin{split}
& \partial_t \langle \varphi \rangle +\langle \varphi \rangle = \langle \uppi \rangle \, , \\
& \partial_t \langle \uppi \rangle +\langle \uppi \rangle = \partial_j \partial^j \langle \varphi \rangle \, . \\
\end{split}
\end{equation}
Thus, we have the dynamical operator:
\begin{equation}
    \Ma(\partial_j) = 
    \begin{bmatrix}
        1 & -1 \\
        -\partial_j \partial^j & 1 \\
    \end{bmatrix} \, .
\end{equation}
The information current $E^\mu$ and the entropy production rate $\sigma$ are
\begin{equation}\label{INFOONA2}
\begin{split}
E^0 ={}& \dfrac{1}{2} \big[ \uppi^2+ \partial_j \varphi \partial^j \varphi \big] \, ,\\
E^j ={}& - \uppi \, \partial^j \varphi \, ,\\
\sigma ={}&  \uppi^2 + \partial_j \varphi \partial^j \varphi \, . \\
\end{split}
\end{equation}
In fact, the dissipation equation $\partial_\mu E^\mu = -\sigma$ holds along all exact solutions of the classical equations of motion \eqref{equiboo}. Furthermore, $E^\mu$ is future-directed non-spacelike, and $\sigma$ is non-negative definite. Hence, both the classical and the fluctuating theory are mathematically consistent, causal, and stable. From \eqref{Ewow}, we obtain the information kernel,
\begin{equation}
    \Ka(\partial_j)= \begin{bmatrix}
        -\partial_j \partial^j & 0 \\
        0 & 1 \\
    \end{bmatrix} \, ,
\end{equation}
so that, recalling equation \eqref{short}, the equal-time correlators immediately follow:
\begin{equation}
    \begin{split}
 \langle \varphi(\x)\varphi(\y)\rangle ={}& \dfrac{1}{4\pi |\x -\y|}  \, , \\
 \langle \varphi(\x)\uppi(\y)\rangle ={}&  0 \, , \\
 \langle \uppi(\x)\uppi(\y)\rangle ={}& \delta^3(\x-\y)  \, . \\
    \end{split}
\end{equation}

\subsection{Forward correlator}

We are now in the position to compute the forward correlators of the present theory. From \eqref{retuzinfinie}, we get
\begin{equation}\label{cacioppoli2}
  \Ga^+(x) =\Theta(t) e^{-t} \int \dfrac{d^3 k}{(2\pi)^3} e^{i\textbf{k}\cdot \x} \,
  \begin{bmatrix}
      \dfrac{\cos(|\kb|t)}{\kb^2} & \dfrac{\sin(|\kb|t)}{|\kb|} \\
      -\dfrac{\sin(|\kb|t)}{|\kb|} & \cos(|\kb|t) \\
  \end{bmatrix} \, .
\end{equation}
The Fourier integral can be solved analytically. In particular, for $t>0$, we obtain
\begin{equation}
\begin{bmatrix}
        \langle \varphi(x)\varphi(0)\rangle & \langle \varphi(x)\uppi(0) \rangle  \\
    \langle \uppi(x)\varphi(0)\rangle & \langle \uppi(x)\uppi(0)\rangle \\
\end{bmatrix}
=
\dfrac{e^{-t}}{4\pi |\x|}
\begin{bmatrix}
    \Theta(|\x|-t) & \delta(|\x|-t) \\
    -\delta(|\x|-t) & -\delta'(|\x|-t) \\
\end{bmatrix} \, .
\end{equation}
As one would expect, the $\uppi \uppi$ (i.e. density-density) correlator propagates on the surface of the future lightcone. On the other hand the $\varphi\varphi$ (i.e. chemical potential-chemical potential) correlator is supported \textit{outside} the future lightcone.

\section{Causal viscosity}\label{causalviscone}

Let us finally study an example of causal and stable viscosity at zero chemical potential.

\subsection{Outline of the theory}

We consider the linearized dynamics of an energy-momentum tensor $\delta T^{\mu \nu}$ with a viscous correction in the form of a shear-stress tensor, and a bulk viscous term. The BDNK degrees of freedom are chosen to be $\Psi=\{\delta T, \delta u^j, \delta v^j \}$, representing respectively the temperature perturbation, the flow velocity, and the momentum per unit enthalpy. The first and the last quantity are taken to be hydrodynamic-frame invariants, defined as follows: $\delta T^{00}=c_v \delta T$, and $\delta T^{0j}=(\epsilon+P)\delta v^j$, where $c_v$ is the specific heat and $\epsilon+P$ is the enthalpy density. The velocity $\delta u^j$, on the other hand, is an infrared effective field, which loses any physical meaning above a certain cutoff scale $\tau$. When the shear and bulk viscosity coefficients are related by the identity $\zeta = 2\eta / 3$, the contributions from shear and bulk combine into a single symmetric tensor, and the averaged equations of motion can be written in the form\footnote{We choose to set $\zeta=2\eta/3$ because this considerably simplifies all later calculations, and makes our analysis easier to follow. For example, it allows us to derive a relatively simple effective action, see equation \eqref{effezio}. On the other hand, we stress that there is no fundamental obstruction in carrying out the construction in the general
case. The information current exists for arbitrary values of $\zeta$ \cite{Gavassino:2024ufs}.} 
\begin{equation} \label{Eq:shear_model}
    \begin{split}
        & c_v \partial_t \langle \delta T\rangle + (\epsilon + P) \partial_j \langle \delta v^j \rangle = 0 \, , \\
        & \tau \partial_t \langle\delta u_j \rangle + \langle \delta  u_j\rangle - \langle \delta v_j \rangle =0 \, ,\\
        & (\epsilon + P) \partial_t \langle \delta v_j \rangle + s \partial_j \langle \delta T \rangle - 2\eta \partial^k \partial_{(k} \langle \delta u_{j)} \rangle = 0 \, .\\
    \end{split}
\end{equation}
This describes the dynamics of a BDNK model for viscous fluid dynamics. The dynamical operator is then 
\begin{equation}
    \Ma (\partial_j) = \begin{bmatrix}
        0 & 0 & (\epsilon + P)/c_v \partial_j \\
        0 & 1/\tau & -1/\tau \\
        s/(\epsilon + P) \partial^k & -\eta / (\epsilon + P) (\delta^k_j \partial^l \partial_l + \partial^k \partial_j) & 0
    \end{bmatrix} \, .
\end{equation}
It can be shown \cite{Gavassino:2024ufs} that all the solutions of this first-order model for viscous hydrodynamics are also exact solutions of a corresponding Israel-Stewart model (but the reversal is not true). This one-way mapping makes it possible to not only prove the causality and stability of the system in equation \,\eqref{Eq:shear_model}, but also derive the exact information current and entropy production rate:
\begin{equation}
\begin{split}
    & T E^0 = \frac{1}{2} \left[ \frac{c_v}{T} \delta T^2 + (\epsilon + P) \delta v_j \delta v^j + 2\eta \tau \partial_{(j} \delta u_{k)} \partial^{(j} \delta u^{k)} \right] \, , \\
    & T E^j = s \delta T \delta v^j - 2\eta \delta v_k \partial^{(k} \delta u^{j)} \, , \\
    & T \sigma = 2 \eta \partial_{(j} \delta u_{k)} \partial^{(j} \delta u^{k)} \, .
\end{split}
\end{equation}
As usual, the dissipation equation $\partial_{\mu} E^{\mu} = -\sigma$ holds along solutions of the equations of motion, and the conditions for causality and stability can be determined \cite{Gavassino:2024ufs}. The information kernel is then 
\begin{equation}
    \Ka (\partial_j) = \dfrac{1}{T} \begin{bmatrix}
        c_v / T & 0 & 0 \\
        0 & -\eta \tau (\delta^k_j \partial_l \partial^l + \partial_j \partial^k) & 0 \\
        0 & 0 & (\epsilon + P) \delta^k_j
    \end{bmatrix} \, .
\end{equation}

\subsection{Correlators of the theory}

The equal time correlators for this theory are presented in \cite{Gavassino:2024ufs}. The symmetric correlators follow from equation \eqref{twoPoint}, and they can be expressed as a block matrix:
\begin{equation}
\begin{bmatrix}
\langle  \delta T(x) \delta T(y) \rangle & \langle \delta T(x)\delta u_j(y) \rangle & \langle \delta T(x) \delta v_j(y)  \rangle \\
\langle \delta u^k(x)  \delta T(y) \rangle & \langle \delta u^k(x) \delta u_j(y)\rangle & \langle \delta u^k(x) \delta v_j(y) \rangle \\
\langle \delta v^k(x)  \delta T(y) \rangle & \langle \delta v^k(x) \delta u_j(y) \rangle & \langle \delta v^k(x) \delta v_j(y) \rangle \\
\end{bmatrix} = \int \dfrac{d^4 k}{(2\pi)^4}e^{ik(x-y)} 
\begin{bmatrix}
\Ga^S_{TT}(k) & \Ga^S_{Tu_j}(k) & \Ga^S_{Tv_j}(k) \\
\Ga^S_{u^k T}(k) & \Ga^S_{u^k u_j}(k) & \Ga^S_{u^k v_j}(k) \\
\Ga^S_{v^k T}(k) & \Ga^S_{v^k u_j}(k) & \Ga^S_{v^k v_j}(k) \\
\end{bmatrix} \, ,
\end{equation}
with 
\begin{equation}
\begin{split}
    \Ga^S_{TT}(k) ={}& \frac{4 \eta T^3 (\epsilon + p)^2 T^4}{|(\epsilon + P)^2 \eta^2 (\tau \omega - i) + T \omega \left[ 2 \eta \textbf{k}^2 - (\epsilon + P) \omega (\tau \omega - i) \right]|^2} \, , \\
    \Ga^S_{Tu_j}(k) ={}& \frac{2 i T^2 (\epsilon + P)^2 k_j \left( (\epsilon + P) \textbf{k}^2 - c_v T \omega^2 \right)}{|(\epsilon + P)^2 \eta^2 (\tau \omega - i) + T \omega \left[ 2 \eta \textbf{k}^2 - (\epsilon + P) \omega (\tau \omega - i) \right]|^2} \, , \\
    \Ga^S_{Tv_j}(k) ={}& \frac{4 \eta c_v T^3 (\epsilon + P) \omega \textbf{k}^2 k_j}{|(\epsilon + P)^2 \eta^2 (\tau \omega - i) + T \omega \left[ 2 \eta \textbf{k}^2 - (\epsilon + P) \omega (\tau \omega - i) \right]|^2} \, , \\
    \Ga^S_{u^k u_j}(k) ={}& \frac{T (\epsilon + P)^2 \left( (\epsilon + P) \textbf{k}^2 - c_v T \omega^2 \right) \left( k^k k_j / \textbf{k}^2 \right)}{|(\epsilon + P)^2 \eta^2 (\tau \omega - i) + T \omega \left[ 2 \eta \textbf{k}^2 - (\epsilon + P) \omega (\tau \omega - i) \right]|^2} + \frac{\frac{2T}{\eta \textbf{k}^2} (\epsilon + P)^2 \omega^2 \left( \delta^k_j - \frac{k^k k_j}{\textbf{k}^2} \right)}{(\epsilon + P)^2 \omega^2 + \left( \eta \textbf{k}^2 - \tau (\epsilon + P) \omega^2 \right)^2} \, , \\
    \Ga^S_{u^k v_j}(k) ={}& \frac{2 i c_v T^2 (\epsilon + P) \omega \left( (\epsilon + P) \textbf{k}^2 - c_v T \omega^2 \right) \left( k^k k_j / \textbf{k}^2 \right)}{|(\epsilon + P)^2 \eta^2 (\tau \omega - i) + T \omega \left[ 2 \eta \textbf{k}^2 - (\epsilon + P) \omega (\tau \omega - i) \right]|^2} + \frac{2iT (\epsilon + P) \omega \left( \delta^k_j - \frac{k^k k_j}{\textbf{k}^2} \right)}{(\epsilon + P)^2 \omega^2 + \left( \eta \textbf{k}^2 - \tau (\epsilon + P) \omega^2 \right)^2} \, , \\
    \Ga^S_{v^k v_j}(k) ={}& \frac{4 \eta c_v^2 T^3 \omega^2 k^k k_j}{|(\epsilon + P)^2 \eta^2 (\tau \omega - i) + T \omega \left[ 2 \eta \textbf{k}^2 - (\epsilon + P) \omega (\tau \omega - i) \right]|^2} + \frac{4\eta T \textbf{k}^2 \left( \delta^k_j - \frac{k^k k_j}{\textbf{k}^2} \right)}{(\epsilon + P)^2 \omega^2 + \left( \eta \textbf{k}^2 - \tau (\epsilon + P) \omega^2 \right)^2} \, .
\end{split}
\end{equation}
The remaining follow immediately from the Hermiticity of $\Ga^S(k)$.
The correlator $\langle \delta v^k(x) \delta v_j(y) \rangle$
corresponds to the physical correlations of the fluid velocity, and it reduces to the standard Navier-Stokes result in the limit $\tau \rightarrow 0$. 

The dual Israel-Stewart model has ten degrees of freedom: the temperature $T$, Landau's flow velocity $v_j$, and the stress tensor $\Pi_{jk}$. The fluctuations of $T$ and $v_j$ in the Israel-Stewart model are equivalent to the fluctuations of $T$ and $v_j$ in the BDNK model above. The symmetric tensor $\Pi_{jk}$ however corresponds to $-2\eta \partial_j u_k$ within BDNK. Since the vector $u_j$ only has three degrees of freedom, while $\Pi_{jk}$ has six, the fluctuations of $\Pi_{jk}$ contain information that is not present in BDNK. This explains why $v_j$, not $u_j$, should be understood as the fluid velocity. The vector $u_j$ is only a proxy that is used to describe the dynamics of the shear stress tensor without including its full degrees of freedom. 

\subsection{Effective action}

The effective action for the fluctuating viscous hydrodynamic model presented in this section can be determined from equations \eqref{effectiveaction} and \eqref{fbruf}. We find that
\begin{equation}\label{effezio}
\begin{split}
    \mathcal{S} = & \int d^4x \, \Bigg[ \delta \tilde{T} \left( \partial_t \delta T {+} \frac{\epsilon {+} P}{c_V} \partial_j \delta v^j \right) + \delta \tilde{u}_j \left( \partial_t \delta u^j {+} \dfrac{\delta u^j{-}\delta v^j}{\tau} \right) + \delta \tilde{v}_j \bigg( \partial_t \delta v^j {+} \frac{\partial^j \delta T}{T}  {-} \frac{\eta}{\epsilon + P} \partial_k\left( \partial^k  \delta u^j {+} \partial^j \delta u^k \right) \bigg) \Bigg] \\
    +& \frac{i T}{\eta \tau^2} \int  d^4x \, d^4y \, \delta (x^0 {-} y^0) \,  \delta \tilde{u}_j(x)  \left( \dfrac{3 \delta^{jk}}{16\pi |\textbf{x}-\textbf{y}|}  + \frac{(x-y)^j (x-y)^k}{16\pi |\textbf{x}-\textbf{y}|^3} \right) \delta \tilde{u}_k(y) .
\end{split}
\end{equation}
The first integral is the ``on-shell" action that encapsulates the equations of motion in the absence of stochastic fluctuations. The second integral expresses the effect of thermal fluctuations on the system, containing the noise correlator. Note that here, the noise term depends only on $\delta u^j$; this is expected as only $\delta u^j$ appears in the entropy production, which is related to the magnitude of fluctuations through the fluctuation-dissipation theorem. 

To evaluate the noise kernel \eqref{fbruf} explicitly, we made use of the following distributional identities:
\begin{equation}
\begin{split}
 \int \dfrac{d^3 k}{(2\pi)^3} e^{i\textbf{k}\cdot \textbf{r}} \, \dfrac{1}{k^4} ={}&  -\dfrac{r}{8\pi} + \text{``Infinite constant''} \, , \\  
\int \dfrac{d^3 k}{(2\pi)^3} e^{i\textbf{k}\cdot \textbf{r}} \, \dfrac{\textbf{k}\textbf{k}^T}{k^4} ={}& \dfrac{1}{8\pi r} \bigg[ \mathbb{I}-\dfrac{\textbf{r}\textbf{r}^T}{r^2} \bigg] \, , \\  
\int \dfrac{d^3 k}{(2\pi)^3} e^{i\textbf{k}\cdot \textbf{r}} \, \dfrac{1}{k^2} ={}& \dfrac{1}{4\pi r} \, , \\
\end{split}
\end{equation}
The first can be verified by solving the integral in spherical coordinates. To obtain a meaningful result, one needs to regularize the denominator, e.g. by replacing $k^4$ with $k^4{+}\varepsilon^4$, and then sending $\varepsilon$ to zero. When this is done, there is a contribution proportional to $\varepsilon^{-1}$ that does not depend on $r$, giving the infinite constant. The second integral can be obtained by differentiating the first twice. The third integral is just the trace of the second.

\section{Origin of long range correlations in BDNK}\label{reparam}

It can be verified that, in the explicit examples considered in this manuscript, the retarded correlator $\Ga^R(x)$ of the primary BDNK fields (temperature, chemical potential, and flow velocity) exits the future lightcone. For example, in the diffusion model of section \ref{nondiffidate}, the retarded correlator takes the form
\begin{equation}
    \Ga^R(x) = \dfrac{\Theta(t)}{T\tau} \int \dfrac{d^3 k}{(2\pi)^3} e^{i\textbf{k}\cdot \textbf{x}} \, e^{-t\Ma(ik_j)}
    \begin{bmatrix}
        (\tau D \textbf{k}^2)^{-1} & -1 \\
        1 & 0 \\
    \end{bmatrix} \, .
\end{equation}
For positive times, this is a classical solution of the equations of motion \eqref{classuzzuz}, whose initial data has infinite support:
\begin{equation}
    \Ga^R(0^+,\textbf{x})=  \dfrac{1}{T\tau} \begin{bmatrix}
        \dfrac{1}{4\pi \tau D |\textbf{x}|} & & -\delta^3(\textbf{x}) \\
        \delta^3(\textbf{x}) & & 0 \\ 
    \end{bmatrix}\, .
\end{equation}
Since the non-locality comes from the $\varphi \varphi$ component, we conclude that the field $\varphi$, which may be interpreted as the chemical potential (in the hydrodynamic frame defined by $\tau$), is not a genuinely local field \cite{HellerBounds2022}, while $\uppi$ is necessarily local (being proportional to a conserved density). Below, we provide a simple explanation for the non-locality of $\varphi$.

\subsection{From Israel-Stewart to BDNK}\label{Istobdnk!!}

Consider an Israel-Stewart theory \cite{Israel_Stewart_1979} of diffusion, with degrees of freedom $\Psi=\{\uppi,q^j \}$, representing respectively a conserved density and its associated flux. Postulate the following (Cattaneo-type \cite{cattaneo1958}) classical equations of motion:
\begin{equation}\label{classuzzuz97}
    \begin{split}
& \tau \partial_t \langle q^j \rangle +\langle q^j \rangle = -D \partial^j \langle \uppi \rangle \, , \\
& \partial_t \langle \uppi \rangle + \partial_j  \langle q^j \rangle=0 \, . \\
    \end{split}
\end{equation}
Then, apply the Helmholtz decomposition theorem, according to which any vector field $q^j$ can be decomposed as the sum of an irrotational vector field and a solenoidal vector field, namely $q^j=-D\big[\partial^j \varphi-(\nabla {\times} A)^j\big]$, with 
\begin{equation}\label{qqqqqqqqqqqq}
\begin{split}
 \varphi(\textbf{x}) ={}& \dfrac{1}{4\pi D}  \int d^3 y \, \dfrac{(\partial_j q^j)_{\textbf{y}}}{|\textbf{x}-\textbf{y}|} \, , \\
 A^j(\textbf{x}) ={}& \dfrac{1}{4\pi D}  \int d^3 y \, \dfrac{(\nabla {\times} q)^j_{\textbf{y}}}{|\textbf{x}-\textbf{y}|} \, . \\
\end{split}
\end{equation}
Now, if one takes the statistical average of the first definition above (the one of $\varphi$), and applies $\tau \partial_t +1$ on both sides, one obtains $\tau \partial_t \langle \varphi \rangle +\langle \varphi \rangle=\langle \uppi \rangle$, which is the first equation of \eqref{classuzzuz}. Furthermore, since $\partial_j q^j=-D\partial_j \partial^j \varphi$, the second equation of \eqref{classuzzuz97} is equivalent to $\partial_t \langle \uppi \rangle =D \partial_j \partial^j \langle \varphi \rangle$, which is the second equation of \eqref{classuzzuz}. Hence, we have recovered the BDNK model for relativistic diffusion discussed in section \ref{nondiffidate}.

The above mapping between classical BDNK and classical Israel-Stewart is exact, at least in the linear regime. It tells us that a BDNK model for diffusion can also be viewed as an Israel-Stewart model for diffusion, provided that we interpret the BDNK chemical potential $\varphi$ as the potential of the irrotational part of the Israel-Stewart flux $q^j$, and the cutoff scale $\tau$ as the Cattaneo-type relaxation time. Similar reasoning was used in \cite{Gavassino:2024ufs} to map model \eqref{Eq:shear_model} into Israel-Stewart's theory for viscosity.

\subsection{Non-locality of the BDNK primary fields}

Let us use the ``duality'' between Israel-Stewart and BDNK in the linear regime to explain the origin of the long-range correlations in $\varphi$. 
The Israel-Stewart degrees of freedom $\uppi$ and $q^j$ in \eqref{classuzzuz97} are \textit{local} physical observables, representing respectively the density and flux of conserved charge. Thus, their correlation is necessarily short-range \cite{landau9}. Indeed, correlators in any Israel-Stewart theory are always supported inside the lightcone \cite{Mullins:2023tjg}. For model \eqref{classuzzuz97}, we have
\begin{equation}
    \langle q^j(\textbf{x})q^k(\textbf{y})\rangle = \dfrac{D}{\tau} \delta^{jk} \delta^3(\textbf{x}-\textbf{y}) \, ,
\end{equation}
at equal times. On the other hand, the BDNK chemical potential $\varphi$, as defined in \eqref{qqqqqqqqqqqq}, is a non-local functional of $q^j$. This intrinsic non-locality is ultimately what generates long-range correlations in $\varphi$. Indeed, if we multiply the first equation of \eqref{qqqqqqqqqqqq} by itself, and take the average, we recover the (long-range) BDNK correlator \eqref{grzuzber}:
\begin{equation}
  \langle \varphi(\x)\varphi(\y)\rangle = \dfrac{1}{4\pi \tau D |\x -\y|}  \, .
\end{equation}

\subsection{Deriving the BDNK action from the Israel-Stewart action}

The MSR action of all Israel-Stewart theories can be straightforwardly constructed following the method of \cite{Mullins:2023ott}. For the diffusion model \eqref{classical}, we have
\begin{equation}\label{Ismsractione}
\mathcal{S}= \int d^4 x \bigg[\tilde{\uppi}(\partial_t \uppi +\partial_j q^j)+\dfrac{\tilde{q}^j}{D}(\tau \partial_t q_j +q_j+D\partial_j \uppi) + i \dfrac{\tilde{q}^j \tilde q_j}{D} \bigg] \, .
\end{equation}
Clearly, this is a perfectly local action. Let us now apply the change of variables introduced in section \ref{Istobdnk!!}, namely $q_j =-D\big[\partial_j \varphi -(\nabla {\times} A)_j\big]$. Integrating by parts, we obtain
\begin{equation}
\mathcal{S}= \int d^4 x \bigg[\tilde{\uppi}(\partial_t \uppi -D\partial_j \partial^j \varphi)+\partial_j\tilde{q}^j(\tau \partial_t \varphi +\varphi- \uppi)+(\nabla {\times} \tilde{q})^j(\tau \partial_t{+}1)(\nabla {\times} A)_j + i \dfrac{\tilde{q}^j \tilde q_j}{D} \bigg] \, . 
\end{equation}
Let us additionally introduce the fields $\tilde{\varphi}=\tau \partial_j \tilde{q}^j$ and $\tilde{A}^j=\tau (\nabla {\times} \tilde q)^j$. Applying the Helmholtz decomposition theorem to vector field $\tilde{q}^j$, we find that
\begin{equation}
4\pi \tau \, \tilde{\textbf{q}}(\textbf{x})= -\nabla \int d^3 y \, \dfrac{\tilde{\varphi}(\textbf{y})}{|\textbf{x}-\textbf{y}|} +\nabla {\times} \int d^3 y \, \dfrac{\tilde{\textbf{A}}(\textbf{y})}{|\textbf{x}-\textbf{y}|} \, ,
\end{equation}
where $\nabla$ denotes differentiation with respect to $\textbf{x}$. With the aid of this formula, we can rewrite $\int d^3 x \, \tilde{q}^j \tilde{q}_j$ as a bilocal integral involving only $\tilde{\varphi}$ and $\tilde{A}^j$. The resulting effective action is reported below:
\begin{equation}
 \mathcal{S}= \int d^4 x \bigg[\tilde{\uppi}(\partial_t \uppi -D\partial_j \partial^j \varphi)+\dfrac{\tilde{\varphi}}{\tau}(\tau \partial_t \varphi +\varphi- \uppi) \bigg]+ i\int \dfrac{d^4 x d^4y}{4\pi D \tau^2} \delta(x^0{-}y^0) \dfrac{\tilde{\varphi}(x)\tilde{\varphi}(y)}{|\textbf{x}-\textbf{y}|} + \text{``action for $\{A^j,\tilde{A}^j\}$''} \, .   
\end{equation}
This shows that the Israel-Stewart action \eqref{Ismsractione}, expressed in terms of $\varphi$ and $\uppi$, coincides with the BDNK action, see equation \eqref{lagringa}. Thus, Israel-Stewart and BDNK are exactly the same stochastic theory\footnote{The change of variables does not affect the measure of the path integral. In fact, the transformation $\{q^j,\tilde{q}^j\} \leftrightarrow \{\varphi,A^j,\tilde{\varphi},\tilde{A}^j \}$ is linear, and its Jacobian determinant is an overall constant that cancels out when we take the quotient on the right-hand side of \eqref{pathintegral}.}. The only difference is that Israel-Stewart possesses an additional dynamical field $A^j$, whose evolution decouples from that of $\{\varphi,\uppi \}$. In particular, $A^j$ does not affect the dynamics of the conserved density. This explains why all the physically meaningful observables of BDNK are indistinguishable from those of Israel-Stewart, in the linear case. It also shows that the apparent non-locality of BDNK is a mathematical artifact of the hydrodynamic frame choice, with no physical relevance.


It should be possible to perform a similar mapping between the BDNK model outlined in section \ref{causalviscone} and the Israel-Stewart theory for viscosity in the Landau frame \cite{OlsonLifsh1990}. The main difficulty in this case is that the analog of the Helmholtz decomposition for the shear stress tensor $\Pi_{jk}$ is far from trivial.

\section{Conclusions}

We have successfully constructed, in the linear regime, a fluctuating formulation of BDNK hydrodynamics that is causal, stable, and mathematically consistent also off-shell. The Martin-Siggia-Rose path integral is always convergent, and the effective action $\mathcal{S}$ obeys the self-consistency requirement $\mathfrak{Im}\mathcal{S}\geq 0$ by construction, healing the pathologies of the previous formulations. Furthermore, the equal-time correlators are consistent with the probability distribution $e^S$ up to first order in derivatives (since the information current $E^\mu$ is exact to first order in the gradient expansion \cite{Gavassino:2024ufs}). Finally, the cutoff scale $\tau$ at which the fluctuating theory breaks down is the same scale $\tau$ at which the deterministic theory breaks down, meaning that the regime of applicability of the stochastic theory is the largest possible.

Unfortunately, the mathematical rigor comes at the expense of physical intuition and aesthetical simplicity. In fact, enforcing both causality and convergence of the MSR path integral forced us to employ elaborate UV-regularization schemes, which gave rise to conspicuous cutoff phenomena. Such phenomena include long-range correlations of the primary fluid variables (such as temperature $T$, chemical potential $\mu$, and flow velocity $u^\mu$), bilocal contributions to $\mathcal{S}$, and an unusual statistical decoupling of conjugate variables at equal times, e.g. $\langle \delta n(\textbf{x}) \delta \mu(\textbf{y}) \rangle=0$. All these effects are artifacts of the regularization scheme, and they arise due to the presence of spurious non-hydrodynamic modes, which cause the effective fields to undergo fast, but long-range, fluctuations at fixed conserved densities. As expected, all the unphysical UV phenomenology cancels out when we examine the transport of conserved charges below the cutoff frequency scale $\tau^{-1}$, giving sensible results for all observable quantities. 

Perhaps the most surprising result of our analysis is that (in the linear regime) the two-point correlators of the conserved densities, such as $\langle \delta T^{0\nu}(x)\delta T^{0\rho}(y)\rangle$, are indistinguishable from those of the Israel-Stewart theory \cite{Mullins:2023tjg} \textit{at all scales} (even beyond the respective regimes of applicability), provided that we identify the BDNK cutoff scale $\tau$ with the Israel-Stewart relaxation time \cite{Israel_Stewart_1979}.
This result is far from obvious, for two reasons. First, BDNK theory and Israel-Stewart theory are profoundly different field theories, with different choices of dynamical degrees of freedom. Secondly, and more importantly, because all the field-field correlators of the Israel-Stewart theory are short-range, and the MSR action is fully local, while fluctuating BDNK exhibits strong non-localities (despite being causal). 

To better understand this unexpected correspondence, we studied the relationship between Israel-Stewart and BDNK frameworks in the case of a simple model for causal diffusion (in the linear regime), uncovering the following mechanism. The degrees of freedom $\Psi_{IS}$ of the Israel-Stewart theory can be  decomposed into a ``BDNK part'' $\Psi_{BDNK}$ and purely transient part $\Psi_{TR}$, such that the effective MSR action splits \textit{exactly} into two disconnected parts:
\begin{equation}
    \mathcal{S}_{IS}[\Psi_{IS}]=\mathcal{S}_{BDNK}[\Psi_{BDNK}]+\mathcal{S}_{TR}[\Psi_{TR}] \, .
\end{equation}
This makes Israel-Stewart and BDNK physically equivalent (in the linear regime), since the dynamics of $\Psi_{TR}$ fully decouples from the dynamics of the conserved quantities. Furthermore, we found that $\Psi_{BDNK}$ is a non-local functional of $\Psi_{IS}$, so the non-locality of linearised BDNK is a direct consequence of the locality of Israel-Stewart, and it has no physical impact on causality. 
Intuitively, the decomposition of $\Psi_{IS}$ into $\Psi_{BDNK}$ and $\Psi_{TR}$ is similar to the decomposition of the Klein-Gordon equation,
\begin{equation}
    -\partial_t^2 \phi = (m^2-\partial^2_x)\phi \, ,
\end{equation}
into a particle and an antiparticle sector:
\begin{equation}
\begin{split}
i\partial_t \phi_+={}& +\sqrt{m^2-\partial^2_x} 
 \, \phi_+ \, ,\\
i\partial_t \phi_-={}& -\sqrt{m^2-\partial^2_x} \, \phi_- \, .\\
\end{split}
\end{equation}
Individually, the two parts $\phi_+$ and $\phi_-$ undergo non-local dynamics, because they themselves are non-local functionals of $\phi$ \cite{BaratLocalization2003,GavassinoDispersion2023}. Still, they are separate degrees of freedom of a perfectly local theory. In a similar way, linearised stochastic BDNK appears to be a non-local field theory, but it is ultimately a ``sector'' of linearised stochastic Israel-Stewart theory, which is manifestly local.


In fact, both Israel-Stewart and BDNK lead to exactly the same physical predictions (in the linear regime), with the difference that:
\begin{itemize}
\item Adding fluctuations to the Israel-Stewart theory in both special and in general relativity can be done through systematic techniques, since it is possible to calculate $E^\mu$ directly from the constitutive relations \cite{GavassinoGibbs2021}. We do not have a similar technique in BDNK, and the present method does not generalize to curved spacetimes.
\item The Israel-Stewart theory admits a fully non-linear Schwinger-Keldysh formulation \cite{Jain:2023obu}, while BDNK does not seem to have a consistent Schwinger-Keldysh formulation even in the linear regime \cite{Mullins:2023ott}.
\item In the non-linear regime, the primary fluid variables such as the temperature and chemical potential will couple in non-trivial ways. For BDNK, this will lead to interactions between short-range and long-range fluid variables, making it difficult to determine whether the correlation functions of conserved currents will remain short-range.
\item In the local rest frame, the noise in BDNK is correlated in space. Due to the relativity of simultaneity, such noise correlations become non-Markovian in a boosted frame. In other words, the noise depends on the whole spacetime history of the fluid. Such non-Markovian behavior has been seen in microscopic derivations of fluctuating relativistic mechanics \cite{Petrosyan:2021lqi}, but it poses a challenge for potential numerical simulations of non-linear fluctuating BDNK. 
\end{itemize}
Given the above facts, we believe that, if the goal is to implement fluctuations in a relativistically covariant framework, it is perhaps best to stick to the Israel-Stewart theory.

\section*{Acknowledgements}

LG is partially supported by a Vanderbilt's Seeding Success Grant. NM is partially supported by the U.S. Department of Energy, Office of Science, Office for Nuclear Physics
under Award No. DE-SC0023861. 
MH was supported by the National Science Foundation (NSF) within the framework of the MUSES collaboration, under grant number OAC-2103680.
We would like to thank N. Abbasi and A. Zaccone for useful exchanges. We also thank K. Ingles and J. Noronha for reading the manuscript and providing useful comments.


\appendix

\section{Proofs of the Fox-Uhlenbeck FDT}\label{AAA}

Here, we provide two proofs that equation \eqref{eiught} implies equation \eqref{nuine}. Throughout this appendix, the dependences of the matrices on $k_j \in \mathbb{R}^3$ will be understood. This does not cause any harm, because in both equations \eqref{eiught} and \eqref{nuine} the wavevector is fixed.

\subsection{Residue theorem}\label{elgringooo}

Defined the matrix $\La(k_j)=\Sa(ik_j) \Qa(k_j) \Sa^\dagger(ik_j)$, we only need to show that
\begin{equation}\label{A1}
     \int_{-\infty}^{+\infty} \dfrac{d\omega}{2\pi} \big[\Ma{-}i\omega\big]^{-1}\La\big[\Ma^\dagger{+}i\omega\big]^{-1} = \Ka^{-1}    
\end{equation}
implies $\La =\Ma \Ka^{-1}+\Ka^{-1}\Ma^\dagger$. Since both equations are continuous in $\Ma$, it will suffice to prove the result assuming that $\Ma$ is diagonalizable. Then, given that diagonalizable matrices are dense in the space of all matrices, the result will automatically hold for all $\Ma$ by continuity. 

By assumption, we have that $\Ma=\sum_n i\omega_n \Pa_n$, where $i\omega_n$ are eigenvalues of $\Ma$, and $\Pa_n$ are eigenprojectors, with $\sum_n \Pa_n = \Ia$, and $\Pa_m \Pa_n = \delta_{mn}\Pa_n$ \cite{Kato_book}. It immediately follows that $\Pa_n \Ma = \Ma \Pa_n=i\omega_n \Pa_n$. Therefore, if we multiply equation \eqref{A1} on the left by $\Pa_m$ and on the right by $\Pa_n^\dagger$, we obtain
\begin{equation}
    \Pa_m \La \Pa_n^\dagger \int_{-\infty}^{+\infty} \dfrac{d\omega}{2\pi}  \dfrac{1}{(\omega-\omega_m)(\omega-\omega_n^*)} = \Pa_m \Ka^{-1} \Pa_n^\dagger \, .
\end{equation}
The integral can be evaluated using the residue theorem, and the result is $-i(\omega_m-\omega_n^*)^{-1}$. To get this outcome, one needs to keep in mind that the functions $\omega_n(k)$ are just the dispersion relations of the classical theory \eqref{classical}, since the latter are solutions of the equation $\det(\Ma-i\omega)=0$. Hence, by hydrodynamic stability, we know that $\omega_m$ lays in the lower $\omega$-plane, while $\omega_n^*$ lays in the upper $\omega$-plane, meaning that, no matter on which side we close the contour integral, there will always be one and only one residue, giving the same final result. Thus, we have that
\begin{equation}
    \Pa_m \La \Pa_n^\dagger = (i\omega_m-i\omega_n^*) \Pa_m \Ka^{-1} \Pa_n^\dagger \, .
\end{equation}
Summing over all $m$ and $n$, we finally obtain $\La=\Ma \Ka^{-1}+\Ka^{-1}\Ma^\dagger$, which is what we wanted to prove.

\subsection{Matrix exponentials}

We can also provide a second proof. First, let us note that, if the classical theory \eqref{classical} is stable, then $e^{-\Ma t}$ decays exponentially to zero at large $t$. Hence, for $\omega\in \mathbb{R}$, we have the following well-known identities:
\begin{equation}\label{freesbone}
   \begin{split}
\big[\Ma{-}i\omega\big]^{-1} ={}&  \int_0^{+\infty} e^{-(\Ma-i\omega)t}dt \, ,\\
\big[\Ma^\dagger{+}i\omega\big]^{-1} ={}&  \int_0^{+\infty} e^{-(\Ma^\dagger+i\omega)s}ds \, .\\
   \end{split}
\end{equation}
Plugging these formulas into \eqref{A1}, and integrating in $\omega$, we obtain
\begin{equation}
    \int_0^{+\infty} e^{-\Ma t} \mathbb{L}e^{-\Ma^\dagger t} dt = \Ka^{-1} \, .
\end{equation}
Multiplying on the left by $e^{\Ma \epsilon}$ and on the right by $e^{\Ma^\dagger \epsilon}$ (with $\epsilon \in \mathbb{R}$), and performing the change of integration variable $t-\epsilon \rightarrow t$, we obtain
\begin{equation}
    \int_{-\epsilon}^{+\infty} e^{-\Ma t} \mathbb{L}e^{-\Ma^\dagger t} dt = e^{\Ma \epsilon}\Ka^{-1}e^{\Ma^\dagger \epsilon} \, .
\end{equation}
Differentiating both sides with respect to $\epsilon$, and evaluating the result at $\epsilon=0$, we finally arrive at the fluctuation-dissipation theorem: $\La = \Ma \Ka^{-1}+\Ka^{-1}\Ma^\dagger$. This completes our second proof.

\section{Evolution of the forward correlator}\label{freestreaming}

If we evaluate equation \eqref{Langevin} at a point $x=(t,\x)$ with $t>0$, we right-multiply it by $\Psi^T(0)$, and take the statistical average, we obtain a partial differential equation for $\Ga^+(x)=\langle \Psi(x)\Psi(0)\rangle$, which holds only for positive times:
\begin{equation}\label{qwertyuiop}
    \big[\partial_t +\Ma(\partial_j)\big] \Ga^+(x) = \Sa(\partial_j)\langle \xi(x)\Psi^T(0)\rangle  \spc (\text{if }t>0) \, .
\end{equation}
On the other hand, we can also solve equation \eqref{Langevin}, so that we have
\begin{equation}
    \Psi(0)= \int_{-\infty}^0 e^{s\Ma(\partial_j)} \Sa(\partial_j)\xi(s) ds \, .
\end{equation}
This implies that the correlator on the right-hand side of \eqref{qwertyuiop} can be expanded in terms of averages of the form $\langle \xi(t,\textbf{x})\xi^T(s,\textbf{y})\rangle$, where $t>0$ and $s \leq 0$. However, the noise is assumed Markovian (i.e. noises at different times are uncorrelated), so that $\langle \xi(t,\textbf{x})\xi^T(s,\textbf{y})\rangle \propto \delta(t{-}s)=0$, and the right-hand side of \eqref{qwertyuiop} vanishes identically. Thus, $\big[\partial_t +\Ma(\partial_j)\big] \Ga^+=0$ at positive times, which is what we wanted to prove. 



\section{Proofs for MSR}

\subsection{Jacobian determinant}\label{BBB}

The proportionality constant between $\delta^{\infty}\big[\Psi-\Psi[\xi]\big]$ and $\delta^{\infty}\big[\partial_t \Psi {+}\Ma\Psi {-}\Sa\xi\big]$ is a functional Jacobian determinant:
\begin{equation}\label{Jacubs}
    \mathcal{J}= \text{det} \bigg[ \dfrac{\delta (\partial_t \Psi {+}\Ma\Psi {-}\Sa\xi)}{\delta \Psi} \bigg] = \text{det} \bigg[ \dfrac{\delta (\Da \Psi {-}\Sa\xi)}{\delta \Psi} \bigg]  \, ,
\end{equation}
where, to lighten the notation, we introduced the operator $\Da(\partial_\mu)=\Ia \partial_t +\Ma(\partial_j)$. We have that
\begin{equation}
    \Da(\partial_\mu) \Psi(x) = \Da(\partial_\mu) \int d^4 \tilde{x} \, \delta^4(x-\tilde{x}) \Psi(\tilde{x}) \, .
\end{equation}
Therefore, the functional derivative in \eqref{Jacubs} reads explicitly
\begin{equation}
 \dfrac{\delta (\Da \Psi {-}\Sa\xi)_x}{(\delta \Psi)_y} = \Da(\partial_\mu) \delta^4 (x-y) \, , 
\end{equation}
and does not depend on $\Psi$ or $\xi$. Note that, for this result to hold, the linearity of the equation of motion \eqref{Langevin} is crucial. For examples, if there was a term proportional to $\Psi^2$ in \eqref{Langevin}, then the Jacobian $\mathcal{J}$ would depend on $\Psi$.

There is one more subtlety that we need to discuss. We have shown that the proportionality constant between $\delta^{\infty}\big[\Psi-\Psi[\xi]\big]$ and $\delta^{\infty}\big[\partial_t \Psi {+}\Ma\Psi {-}\Sa\xi\big]$ is the determinant of $\partial_t +\Ma$. The careful reader may have noticed that such a determinant may in general happen to be zero, making the whole change of variables in the Dirac delta meaningless. Luckily, this does not happen in dissipative systems. In fact, the kernel of $\partial_t +\Ma$ is the space of solutions of the classical equations of motion. If such equations of motion are dissipative, the associated solutions must decay in time, and they are, therefore, unbounded in the past. For this reason, they are naturally excluded from the measure of the functional integral, making the operator $\partial_t +\Ma$ invertible in the functional space of interest.

\subsection{Evaluation of the stochastic kernel}\label{BBB2}

The effective action $\mathcal{S}$ can be expressed as a generalized quadratic form as follows:
\begin{equation}
    -i\mathcal{S} = \dfrac{1}{2} \int d^4 x \, d^4 y \, \big(\Psi^T(x),\Tilde{\Psi}^T(x) \big) \mathbb{W}(x-y) \begin{pmatrix}
        \Psi(y) \\
        \Tilde{\Psi}(y) \\
    \end{pmatrix} \, ,
\end{equation}
where the kernel $\mathbb{W}$ can is a square block matrix, given by the Fourier integral
\begin{equation}
    \mathbb{W}(x-y)=\int \dfrac{d^4 k}{(2\pi)^4} e^{ik(x-y)} \begin{bmatrix}
        0 & -i\Da^\dagger(ik) \\
        -i\Da(ik) & \Ha(k) \\
    \end{bmatrix} \, ,
\end{equation}
with $\Da(\partial_\mu)$ defined as in appendix \ref{BBB}. It is well known that the Fourier components of propagators of Gaussian path integrals can be obtained by simply inverting the kernel matrix in Fourier space, namely\footnote{Note that $\Da^\dagger(ik)=\Da^T(-ik)$ for all real $k$. To prove this fact, we just need to note that the function $f(\lambda)=e^{-\lambda kx}\Da(\partial)e^{\lambda kx}=\Da(\lambda k)$ is entire for $\lambda \in \mathbb{C}$, and it takes real values for real $\lambda$. Thus, we can apply the Schwartz Reflection Principle \cite{NeedhamVisualComplexAnalysis}, and we have that $[f(\lambda)]^*=f(\lambda^*)$, where $*$ denotes ordinary complex conjugation. Taking $\lambda=i$, we obtain $[\Da(i k)]^*=\Da(-i k)$.}:
\begin{equation}\label{amoinvertiere}
    \begin{bmatrix}
        \langle \Psi(x)\Psi^T(y) \rangle & \langle \Psi(x)\Tilde{\Psi}^T(y) \rangle  \\
        \langle \Tilde{\Psi}(x)\Psi^T(y) \rangle  &  \langle \Tilde{\Psi}(x)\Tilde{\Psi}^T(y) \rangle \\
    \end{bmatrix}=\int \dfrac{d^4 k}{(2\pi)^4} e^{ik(x-y)} \begin{bmatrix}
        0 & -i\Da^\dagger(ik) \\
        -i\Da(ik) & \Ha(k) \\
    \end{bmatrix}^{-1} \, .
\end{equation}
Note that only the block $\langle \Psi(x)\Psi^T(y) \rangle$ in the above matrix is a statistical average in the proper sense. In fact, the weight $e^{i\mathcal{S}}$ of the path integral is in general complex, and it cannot be interpreted as a probability distribution. Thus, the ``average'', say, $\langle \Psi(x)\tilde{\Psi}(y)\rangle$ may be complex-valued, even if both $\Psi$ and $\tilde{\Psi}$ are real-value quantities. On the other hand, evaluating the path integral \eqref{pathintegral} is formally equivalent to averaging $F[\Psi]$ over its probability distribution, as given in \eqref{veciopreston}. For this reason, only the correlator $\langle \Psi(x)\Psi^T(y) \rangle$ is physically meaningful, being a genuine average. 

The block-matrix inverse in \eqref{amoinvertiere} can be evaluated explicitly using Corollary 3.3 of \cite{Lu2002}, giving
\begin{equation}
   \langle \Psi(x)\Psi^T(y) \rangle =  \int \dfrac{d^4 k}{(2\pi)^4} e^{ik(x-y)} \Da(ik)^{-1} \Ha(k) \Da(ik)^{-\dagger} \, .
\end{equation}
Comparing this expression with \eqref{twoPoint}, and considering that $\Da=\Ma-i\omega$, we find that $\Ha {=} \Ma \Ka^{-1}{+}\Ka^{-1}\Ma^\dagger$. Given that $\Ha(k)$ does not depend on $\omega$, we can perform the integral in time, and obtain \eqref{fbruf}.

\bibliography{Biblio}

\label{lastpage}

\end{document}